\documentclass[
    ,final            
    ,draft            
  ,numberedheadings 
  ]
  {aipproc}

\layoutstyle{8x11single}

\usepackage{graphicx}
\usepackage{epsfig}
\usepackage{amssymb,amsmath,latexsym,mathrsfs}

\newcommand{\be}{\begin{equation}}
\newcommand{\ee}{\end{equation}}
\newcommand{\bea}{\begin{eqnarray}}
\newcommand{\eea}{\end{eqnarray}}
\newcommand{\mr}{\mathrm}
\def\nicefrac#1/#2{\leavevmode\kern.1em
\raise.5ex\hbox{\the\scriptfont0 #1}\kern-.1em
/\kern-.15em\lower.25ex\hbox{\the\scriptfont0 #2}}

\newcommand{\bx}{{\mathbf x}}
\newcommand{\bn}{{\mathbf n}}
\newcommand{\cd}{\cdot}
\newcommand{\lan}{\langle}
\newcommand{\ran}{\rangle}
\newcommand{\CC}{{\cal C}}
\newcommand{\BB}{{\cal B}}
\newcommand{\EE}{{\cal E}}
\newcommand{\HH}{{\cal H}}

\newcommand{\OO}{{\cal O}}
\newcommand{\dd}{\partial}
\newcommand{\pa}{{\parallel}}
\newcommand{\ub}{\mathrm{b}}
\newcommand{\ybr}{y_\ub}
\newcommand{\bk}{\mathbf{k}}
\newcommand{\dec}{\mathrm{dec}}
\newcommand{\kappafive}{\kappa_{_5}}
\newcommand{\ra}{\rightarrow}
\newcommand{\ie}{{\em i.e. }}
\newcommand{\eg}{{\em e.g. }}
\newcommand{\da}{\dot{a}}

\newcommand{\al}{\alpha}
\newcommand{\de}{\delta}
\newcommand{\lap}{\Delta}
\newcommand{\De}{\Delta}
\newcommand{\ep}{\epsilon}

\newcommand{\ga}{\gamma}
\newcommand{\Ga}{\Gamma}
\newcommand{\La}{\Lambda}
\newcommand{\la}{\lambda}
\newcommand{\ka}{\kappa}
\newcommand{\om}{\omega}
\newcommand{\Om}{\Omega}
\newcommand{\vph}{\varphi}
\newcommand{\si}{\sigma}
\newcommand{\Si}{\Sigma}
\newcommand{\vth}{\vartheta}
\newcommand{\Up}{\Upsilon}
\newcommand{\ze}{\zeta}

\newcommand{\gsim}{\stackrel{>}{\sim}}
\newcommand{\lsim}{\stackrel{<}{\sim}}
\newcommand{\tension}{\la}
\newcommand{\real}{\ensuremath{\mathbb R}}       
\newcommand{\ve}[1]{\boldsymbol{#1}}           
\newcommand{\babla}{{^b\nabla}}                  

\newcommand{\etal}{\textsl{et al.~}}

\begin{document}

\title{Braneworlds}
\author{Ruth Durrer}{address={
 Universit\'e de Gen\`eve, \'Ecole de Physique, 24 Quai E. Ansermet,
CH1211 Gen\`eve 4, Switzerland}}

\begin{abstract}
This course is an introduction to the physics of braneworlds.
We concentrate on braneworlds with only one extra-dimension and
discuss their gravity. We derive the gravitational equations on the
brane from the bulk Einstein equation and explore some limits in which
they reduce to 4-dimensional Einstein gravity. We indicate how cosmological
perturbations from braneworlds are probably very different from usual
cosmological perturbations and give some examples of the preliminary
results in this active field of research.\\
For completeness, we also present an introduction to 4-dimensional
cosmological perturbation theory and, especially its application to
the anisotropies of the cosmic microwave background. 
\end{abstract}

\maketitle

\section{Introduction}
During recent years, cosmology has become one of the most successful fields
in physics. The precise measurements of the anisotropies in the cosmic
microwave background have confirmed a simple 'concordance model': The
Universe is spatially
flat. Its energy density is dominated by vacuum energy (or a
cosmological constant) which contributes about 70\% to the expansion
of the Universe, $\Omega_\La \simeq 0.7$.
The next important contribution is cold dark matter (CDM) with
$\Omega_{CDM} \simeq 0.3$. The expansion velocity of the Universe is
given by the Hubble constant $H_0 = 100h$km/sMpc, with $h\simeq
0.7$. Baryons only contribute a small portion of $\Omega_bh^2
\simeq 0.02$. Massive neutrinos contribute similarly or less.

The structures in the Universe (galaxies, clusters, voids and
filaments) have formed out of small initial fluctuations which have
been generated during inflation and have an
almost scale-invariant spectrum, $n=1\pm 0.1$. There is probably also
a small amount of tensor fluctuations (gravity waves) generated during
inflation, which however has not yet been detected. 

All the above numbers are accurate to a few
percent and will be measured even more precisely with ongoing and
planned experiments. This
situation is unprecedented in cosmology. About twenty years ago,  these numbers
where known at best within a factor of two or even only by their order
of magnitude.
The concordance model is in agreement with most cosmological data, most notably
the CMB anisotropy measurements, supernova type Ia distances (see contribution
by Varun Shani), statistical analysis of the galaxy distribution,
constraints from cosmic nucleosynthesis, cluster abundance and evolution etc.

However, on a theoretical level our understanding has remained poor. We have no
satisfactory answers to the questions:
\begin{itemize}
\item What is dark matter ?
\item What is dark energy?
\item What is the 'inflaton'? Or what is, more precisely, the physics
  of inflation?  
\item How can we resolve the Big Bang and other singularities of classical
general relativity?
\end{itemize}

There is justified hope that the last question could be resolved within
a theory of quantum gravity, which is anyway needed if we want to put all
fundamental interactions on a common footing. At present, the most successful
attempt towards a theory of quantum gravity is string theory. This theory is
based on the assumption that the 'fundamental objects' are not particles but
one dimensional strings. Particles then manifest as excitations, proper modes,
of strings. It would lead us much too far to give an introduction to
string theory at this point. The interested student will have to study the
two volumes of Polchinski~\cite{Pol}.

In this course we next give an introduction to braneworlds, or, more
generally, to physical effects of extra dimensions. 
In Section~3, we derive the gravitational equations for braneworlds
with one co-dimension from Einstein's equations in the bulk. We
then discuss in detail the Randall--Sundrum II model, its background
and its perturbations. In Section~5, we  give an introduction to
4-dimensional cosmological perturbation theory and, especially to the
CMB anisotropy spectrum. Only after this we are ready for braneworld
cosmology in Section~6. We write down the most general brane
cosmology in an  empty bulk and we  investigate some of its
modifications w.r.t. 4-dimensional cosmology. In particular, we discuss
 the modification of the slow roll parameters during
braneworld inflation. We also present one example of the modifications in the
evolution of cosmological perturbations which are relevant in
braneworlds. We end with some conclusions.
\vspace{0.2cm}\\
{\bf Notation:} We use capital Latin indices $A,B,\cdots$ to denote
bulk coordinates, lower case Greek indices $\mu,\nu,\cdots$ for
coordinates on a four dimensional brane, and lower case Latin indices,
$i,j,\cdots$ for spatial 3-dimensional quantities. We sometimes also
use bold symbols to denote 3d spatial vectors. We use the metric
signature $(-,+,\cdots,+)$. The 4-dimensional Minkowski metric is
denoted by $(\eta_{\mu\nu})$.

Throughout we set $c=\hbar=k_{\rm Boltzmann}=1$ so that time and
length scales are measured in inverse energies (usually GeV's), and mass
and temperature correspond to an energy. The four dimensional Newton
constant is then given by $G_4 =  0.67\times
10^{-38}$GeV$^{-2}$. Useful relations in this set of units are $1=
0.2$GeV fm and 1 eV $= 1.16\times 10^4$K. Here fm $=$ femtometer $=
10^{-15}$m.

\section{Basics of Braneworlds}

\subsection{What are Braneworlds?}
The interest of string theory lies in the fact that it may provide a
unified description of gauge interactions and gravity. Its weak point
is, that it is extremely hard to make  predictions from string theory
which are testable at energies available in experiments. The reason
for that is that string theory probably fully manifests itself only
at very high energies of the order the Planck scale. The observed 4-dimensional
Planck scale is given by Newton's constant, $G_4$. In our units
with $\hbar=c=1$ the Planck scale is $E_4 = M_4 =1/\sqrt{4\pi G_4}  \simeq
3\times 10^{18}$GeV. This energy scale
 cannot be achieved by far at terrestrial accelerators (the LHC presently under
construction at CERN will achieve about 7000GeV).

Nevertheless, string theory makes some relatively firm predictions
which might lead to observational consequences at low energy. First of
all, it predicts that spacetime is ten-dimensional with one time and nine
spatial dimensions. Since the observed world  has only four
 dimensions, one usually assumes that the other six are compact and very
small, so that they cannot be resolved by any physical experiment
available to us so far.

Furthermore, string theory predicts the existence of so called $p$-branes,
$p+1$-dimensional sub-manifolds of the ten dimensional spacetime on which open
strings end. Gauge fields and gauge fermions which correspond to string end
points can only move along these $p$-branes, while gravitons which are
represented by closed strings (loops) can propagate in the full
spacetime, the 'bulk'.

This basic fact of string theory has led to the idea of braneworlds:
it may be that our $3+1$-dimensional spacetime is such a
3-brane. If this is so, only gravity can probe the bulk and the
additional dimensions can be much larger than the smallest length
scale which we have probed so far, which is of the order of
$(200$GeV$)^{-1} \simeq 10^{-18}$m. Actually, Newton's law has been
tested only down to scales of about 0.1mm~\cite{micGra}. Hence, in the
braneworld
picture where only gravity can probe the extra-dimensions, these can
be as large as 0.1mm$=10^{-3}$m. In the next subsection we show how
this fact can be employed to address the hierarchy problem.

\subsection{Lowering the fundamental Planck scale}
The fact that the 4-dimensional Planck scale, $M_4 \sim 10^{19}$GeV
is so much larger than the fundamental scale in elementary particle
physics, the electroweak scale $E_{ew}\simeq 10^3$GeV is called the
hierarchy problem. Apart from it seeming unnatural to have two so
widely separated scales to describe fundamental physics, a more 
serious problem is the fact that as soon as we have a unified quantum
theory which describes also gravity, the scale $M_4$ will enter in
quantum corrections of all electroweak scale quantities which are not
especially protected e.g. by symmetries and it will therefore
completely spoil the so successful low energy standard model.

Here we show that within the braneworld picture, it is possible that
the 4-dimensional Planck scale is not fundamental but only an
effective scale which can become much larger than the fundamental
Planck scale $M_P$ if the extra-dimensions are much large than
$M_P^{-1}$. Our argument goes back to Arkani-Hamed , Dimopoulos and
Dvali (1998)~\cite{ADD}.

Let $M_P$ be the fundamental Planck scale and $L$ the size of
$n$ extra dimensions. In addition there are 3 large
spatial dimensions (and time).  For simplicity we assume the $n$
extra-dimensions to be rolled up as a cylinder, $(S^1)^{\otimes
  n}\otimes \real^3$.
The gravitational constants $G_{(n+4)}$
and $G_4$ are defined by the force laws of gravity. Due to the Gauss
constraint these must have the forms
$$        F_{(n+4)} = G_{(n+4)} \frac{m_1m_2}{r^{n+2}}~,\quad
\mbox{ and } \qquad\quad
F_4 = G_4 \frac{m_1m_2}{r^2}~.    $$

\begin{figure}[ht]
\centerline{\epsfig{figure=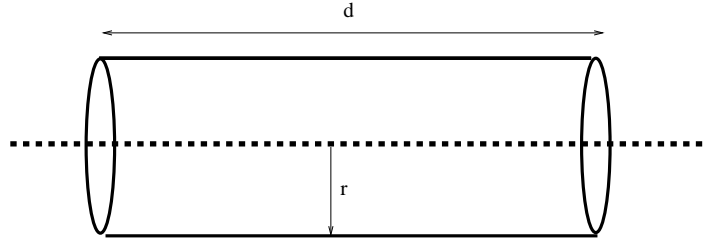, width=9.5cm}}
\caption{A $1+1$ dimensional cylinder around a 1-dimensional grid of
  mass points, corresponding to the de-compactification of a mass point
  on a circle.}
\label{f:cyl}
\end{figure}

On small scales, $r\ll L$, an observer on the brane sees $n+4$
dimensional gravity, while on large scales, $r\gg L$, the cylinder can
no longer be resolved and  simple 4 dimensional gravity is
observed. In order to relate the constants $ G_4$ and $G_{(n+4)} $, we
 de-compactify the compact dimensions leading to an
n-dimensional lattice of masses which looks from far like a
hypersurface of mass density $m/L^n$ (see Fig.~\ref{f:cyl}). Around
the mass distribution, we now form a cylinder $C$ of dimension $n+2$,
length $d$ and radius $r$  (see
Fig.~\ref{f:cyl}). To satisfy Gauss' law,  we require
$$ \int_C F_\perp d\si = S_{(2+n)}G_{(4+n)}\times(\mbox{mass in }C).$$
Here, $F_\perp$ is the component of the force normal to the surface
and $S_j$ is the volume of a $j$ dimensional
sphere, $S_j = 2\pi^{(j+1)/2}/\Gamma ([j+1]/2)$, where $\Gamma$ denotes
the Gamma-function, $\Gamma ([j+1]/2) = ([j-1]/2)!$ for integer values
of $(j+1)/2$.

The first integral is $4\pi r^2d^nF(r)$ while the mass inside
the cylinder is $md^n/L^n$. With the 4-dimensional force law this implies
\be \label{eq:Gn4}
G_4=\frac{S_{(2+n)}}{4\pi}\frac{G_{(4+n)}}{L^n}~.
\ee

In order to relate the gravitational constant to the Planck mass, we express
Newtonian gravity in terms of an action principle. For a static
weak gravitational field $\phi$ and a mass density $\rho$ in $4+n$
dimensions, we can obtain the Poisson equation by varying
the action
$$
I =\int d^{(3+n)}x\left[\frac{M_P^{(2+n)}}{2}\phi\nabla_{(3+n)}^2\phi
    + \rho^{(4+n)}\phi + ...\right]  ~ .
$$
Integrating out $\phi$, we obtain again the Newtonian force law and
the relation
\be \label{e:MPG}
M_P^{(2+n)} = \frac{G_{(4+n)}^{-1}}{S_{(2+n)}}~.
\ee
Integrating the Lagrangian over the extra dimensions relates the 4-
and $4+n$-dimensional Planck masses by
$$  M_4^2 = M_P^{(2+n)}L^n ~.$$
Together  with Eq.~(\ref{e:MPG}) this reproduces
the result (\ref{eq:Gn4}).
For a sufficiently large length scale $L$, $M_4 \simeq 3\times 10^{18}$GeV can
therefore be much larger than the fundamental higher dimensional Planck
scale $M_P$.

Experimental 'micro-gravity' bounds~\cite{micGra} require $L< 0.1$mm.
For $n=2$ and $L\sim 1$mm the fundamental Planck scale can
 be of the order of the electroweak scale,
     $M_P \sim (1 ~ - ~10)$TeV .
This Planck scale seems to be in agreement with most other bounds
(cooling of supernovae, evolution of the Universe, etc) but leads to
the very interesting prospective that effects from string theory might
be observable at the Large Hadron Collider (LHC) presently under
construction at CERN~\cite{bounds}.

Using 'large' extra-dimensions, $L\gg M_P^{-1}$, the fundamental
Planck scale can therefore be of the same order as the
electroweak scale. On the other hand, there is no explanation for the
length scale $L \sim 0.1$mm$\sim 10^{-3}$eV. So the hierarchy problem
has not actually been solved, but it has been moved from an energy
hierarchy to an unexplained length scale. The hope is, that there
would exist solutions of string theory which lead to such a scale
dynamically.

\subsection{New Physics from higher dimensions}
\paragraph{Kaluza-Klein modes}
We denote the four brane coordinates
by $x^\mu$ and the additional $n$ bulk coordinates by $y^a$. For
simplicity, we consider a bulk where the extra dimensions are
rolled up in a cylinder of circumference $L$. In the general situation
where the compact extra dimensions form a non-flat Calabi-Yau manifold,
$\CC$, the exponentials below have to be replaced by the corresponding
eigen-functions of the Laplacian on $\CC$. The case of a warped geometry,
where the extra dimensions may even be non-compact, will be discussed
separately in Section~\ref{sec:RS}.

Be now $\phi$ a massless scalar field in the 'bulk', $\phi(x, y)$ with
$\phi(\cdots,y^a+L\cdots) = \phi(\cdots,y^a,\cdots)$.
    We can expand the y dependence of $\phi$ in Fourier series
      $$ \phi(x, y) = \sum_j \phi_j(x)\exp(i2\pi j\cdot y/L) $$
Since $\phi$ satisfies the massless wave equation,
   $\nabla_{4+n}^2\phi =0$, in the bulk, for a 4d observer which
cannot resolve the scale $L$, the modes $j \neq 0$ will become massive
fields,
$$
(\nabla_4^2+m_j^2)\phi_j = 0  \qquad \mbox{with } \qquad   m_j^2 =
\frac{(2\pi j)^2}{L^2}~.
$$
These modes of the gravitational potential give raise
to exponential corrections to Newton's law,
$$
V(r)  = \frac{G_4}{4\pi r}[1 + \frac{e^{-r/L}}{r^2} + \cdots ]~.
$$

If the extra dimensions are large, the first few masses can be
very low, but if the graviton is the only bulk field, its
weak coupling to other bulk modes leaves the theory
nevertheless viable~\cite{bounds}.

As we shall see in Section~\ref{sec:RS}, the mass spectrum can even be
continuous, $0<m<\infty$, if there are non-compact extra dimensions.

\paragraph{Higher dimensional spin modes}
The spin states of massless particles in $d$ space time dimensions
are characterized by an irreducible representation of $SO(d-2)$.
For $d=4$, massless particles always carry a 1--dimensional
representation of $SO(2)$~\footnote{The full little group for massless
  particles is actually $ISO(2)$, the group of two dimensional
  Euclidean motions and rotations. But for finite dimensional
  representations of $ISO(2)$ the translations are acting
  trivially. The reason why not all
representations of the universal covering group of $SO(2)$,
$\real$, have to be considered, namely $e^{ikx}, ~k\in\real$ is rather
subtle. On the classical level, where $SO(2)$ is the relevant group,
this problem disappears. A full
discussion of the finite dimensional representations of the Poincar\'e
group for massless particles in $d=4$ can be found
e.g. in~\cite{Wein}.}. Taking into account also parity, this
leads to the two helicity modes of all massless particles in
4 dimensions, independent of their spin. This situation changes
drastically if we allow for extra dimensions. Let us consider $d=5$: a massless
particle of spin $s$ in $d=5$ spacetime dimensions, carries the
representation $D^{s}$ of
$SO(3)$ and thus has $2s+1$ spin states, like a massive particle
of spin $s$ in $4$ dimensions.

Let us, for example, consider the
graviton. In $4+1$ dimensions it has the $5$ helicity  states of the
tensor representation of SO(3).
Projected onto a $3+1$ brane, two of them become the usual spin 2
graviton, two are a spin 1 particle, a gravi-vector and one has spin 0,
the gravi-scalar.

The gravi-vector couples to the $\mu 4$ components of the
energy momentum tensor. Interpreting these as the electromagnetic current
$J^\mu$ and the gravi-vector as the electromagnetic potential $A^\mu$,
the five-dimensional Einstein equations lead to Maxwell's equations
for $A^\mu$ and $J^\mu$. This is the so called Kaluza--Klein miracle,
which is also true if any, non--Abelian gauge group replaces the
one-dimensional torus which plays here the role of the electrodynamic
gauge group $U(1)$. This finding of Kaluza and Klein~\cite{KK} has
evoked an interest 
in extra-dimensions in the 20ties, long before string theory.

The positive aspects of the vector sector are, however, over shaded by
the problems coming from the gravi-scalar. It couples to the
four-dimensional energy momentum tensor and modifies gravity. It leads to
a scalar-tensor theory of gravity with several observable consequences.
For example, one can calculate the modification in the slowing down of the
binary pulsar~\cite{bipul} PSR1913+16 due to the radiation of gravi-scalars.
In Ref.~\cite{DK} it is shown that in the simple case of a 5-dimensional
cylindrical bulk, this leads to a modification of the
quadrupole formula by about 20\%, while observations agree with the quadrupole
formula to better than $\frac{1}{2}$\%. If there are more than one
extra-dimensions, there are several gravi-scalars and this problem is
only enhanced.

Clearly, a modification of higher dimensional gravity is necessary
to address
the problem. Usually, one gives a mass to the gravi--scalar. There are several
ways to do this and the resulting four dimensional theory in general depends
on this choice. One proposal is the Goldberger--Wise
mechanism~\cite{GW}.

Another solution is offered by non-compact extra-dimensions. As we
shall see, in curved spacetimes, extra-dimensions can even be
infinite. Due to a so called 'warp factor' they become very small when
seen from the brane. For infinite extra dimensions it can happen that
the gravi-scalar represents a non-normalizable and therefore unphysical mode.
This is precisely what happens in the Randall--Sundrum model and we
shall discuss it in this context in Section~\ref{sec:RS}.

\section{Geometry of five dimensional Braneworld geometry}
From now on we restrict ourselves to five dimensional braneworlds, \ie
braneworlds with only one extra-dimension. The idea here is still that
spacetime has 10 (or for M-theory 11) dimensions, but 5 (or 6) of them
are compactified to a static manifold of about Planck scale, while
one of them is large. This picture is motivated mainly from
11-dimensional M-theory, \eg the Horava-Witten model~\cite{HOWI}, but
we shall not try to implement a realization of this model
here. Nevertheless, it is important to note, that one co-dimension
differs significantly from more than one. The main point is that the
3-brane splits space into two parts, the 'left' and the 'right' hand side
of the brane.  We shall see, that in this case it is possible to
determine the gravitational equations on the brane by simply
postulating Einstein's equations in the bulk. This is no longer
possible in the case of two or more extra-dimensions.

In this section we derive and discuss these brane gravity equations.
 In the next section we shall apply them to the Randall--Sundrum
 model, which we consider as beeing so far the most promising
braneworld model.

To determine the gravitational equations on the brane, we start from the basic
hypothesis that  string theory predicts Einstein gravity in the bulk,
\be \label{eq:Ein}
   G_{AB} = \kappa_5 T_{AB}~.
\ee
We want to discuss in detail the situation of a $3$- brane in a
5--dimensional bulk. We denote the bulk coordinates by
$(x^A)=(x^\mu,y)$, where $(x^\mu)$ are coordinates along the brane and
$y$ is a transverse coordinate. We denote the brane position by
$y=y_b$; in general $y_b$ depends on the point on the brane, $y_b=y_b(x^\mu)$.
A more general embedding of the brane will be discussed below.
Very often we consider an energy momentum tensor of the form
\be\label{eq:Tbulk}
T_{AB} ((x^C)=(x^\la,y))= \frac{\Lambda_5}{\ka_5} g_{AB} +\de_A^\mu \de_B^\nu
   T_{\mu\nu}((x^\la))\de(y-y_b)~.
\ee
Here $\La_5$ is a bulk cosmological constant, $T_{\mu\nu}$ is the
energy momentum tensor on the brane and $\ka_5 =6\pi^2G_5$ is the
five-dimensional gravitational coupling constant. This is the most
general ansatz if we do not allow for any matter fields in the bulk.

\subsection{The second fundamental form}

As above, $g_{AB}$ is the bulk metric. We denote  the projection
operator onto the brane
by $q^A_\mu$. The induced metric on the brane, also called the first
fundamental form, is then given by
\be\label{eq:firstfund}
g_{\mu\nu}(x^\la) =q_\mu^A(x^\la)q_\nu^B(x^\la)g_{AB}(x^\la,y_b(x^\la)).
\ee
We denote the covariant derivative in the bulk by $\babla_A$ and
covariant derivative on the brane (with respect to the induced metric)
by $\nabla_\mu$. We also introduce the brane normal $n$, a vector
field defined on the brane which is normal to all vectors parallel to
the brane. Clearly, for an 
arbitrary  vector field $X = X^\mu\dd_\mu$ along the brane,
$\nabla_\mu X \neq \babla_\mu X$. The difference of these two
covariant derivatives is given by the extrinsic curvature also called
the second fundamental form which we now introduce.

Be $X = X^\mu\dd_\mu$ and $Y = Y^\mu\dd_\mu$  two vector fields on the
brane. Their  covariant derivative on the brane is given by
$$
\nabla_YX = (Y^\mu\dd_\mu X^\nu + \Ga_{\mu\beta}^\nu X^\mu Y^\beta)\dd_\nu~,
$$
while the covariant  derivative in the bulk is
$$
\babla_YX = (Y^\mu\dd_\mu X^\nu + \Ga_{\mu\beta}^\nu X^\mu
Y^\beta)\dd_\nu + \Ga_{\mu\beta}^4\dd_4~.
$$
Therefore, there exists a bi-linear form $K_{\mu\nu}$ on the brane
such that
\be \label{e:2nd}
\babla_YX = \nabla_YX +K(X,Y)n~.
\ee
  Since
$\babla_YX - {\babla}_XY =[Y,X]$  is tangent to the brane, $K(X,Y)-K(Y,X)=0$,
hence $K$ is symmetric. $K$ is called the 'second fundamental form' or
 'extrinsic curvature' of the brane. Its sign is not uniquely
defined in the literature; we shall use Eq.~(\ref{e:2nd}) as its definition.

Since $n$ is the unit normal of the  brane, $g(n,\babla_YX) =
K(Y,X)$. But as $n$ is normal to the brane vector field $X$ we have
$ 0 = \babla_Y(g(n,X)) = g(\babla_Yn,X) + g(n,\babla_YX)$,
so that
$$
K(Y,X) = -g(\babla_Yn,X) = -\frac{1}{2}\left[g(\babla_Yn,X) +
g(\babla_Xn,Y) \right]~.
$$
In components, $K(X,Y) = K_{\mu\nu}X^\mu Y^\nu$, this becomes
\be \label{e:2nd2}
K_{\mu\nu} = -\frac{1}{2}\left[\babla_\mu n_\nu + \babla_\nu n_\mu\right]~.
\ee
Close to the brane we can choose coordinates such that
$$
g_{AB}dx^Adx^B \equiv ds_b^2 = g_{\mu\nu}dx^\mu dx^\nu + dy^2~.
$$

In these so called 'Gaussian normal coordinates', $n= \dd_y$ and
$\babla_\mu n_\nu = \Ga^4_{\mu\nu} = -(1/2)g_{\mu\nu},_4$. The second
fundamental form then becomes simply
$$  K_{\mu\nu} = -\frac{1}{2} \dd_yg_{\mu\nu}  ~. $$

For general coordinates $(z^\mu)$ on the brane we have to define  a brane
parameterization $x^A = X_b^A(z^\mu)$. The vector fields
$(e_\mu) = (\dd_\mu X_b^A(z)\dd_A)$ then form a basis of tangent
vectors on the brane. In terms of these one obtains by means of
Eq.~(\ref{e:2nd2})
\be\label{eq:Kmn}
 K_{\mu\nu} = -\frac{1}{2}\left[g_{AB}(e_\mu^A\dd_\nu n^B +
   e_\nu^A\dd_\mu n^B) +   e_\nu^Ae_\mu^Bn^Cg_{AB,C}\right] ~,
\ee
where a comma denotes an ordinary derivative, $f_{,C} =\frac{\dd
  f}{\dd x^C}$. 

\subsection{The junction conditions}
Einstein's equations with a thin hyper-surface of matter become singular
since there is a $\de$- function in the energy momentum tensor.
Integrating them once across the brane leads to the so called
junction or jump conditions  (of Israel, Lancos,
Darmois, Misner)~\cite{Is,La,Da,Mi}.

Before we come to the algebraically more complicated situation of
general relativity, let us first recall the well known junction conditions of
electrostatics: we consider a conducting boundary surface  (e.g. capacitor
plate) with surface charge density $\rho$ and current density $\ve j$
along the surface.

\begin{figure}[ht]
\vspace*{0.3cm}
\centerline{\epsfig{figure=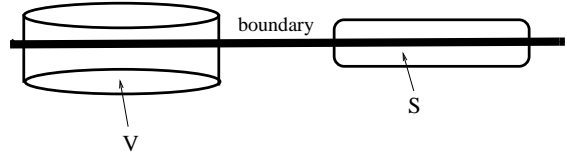, width=2cm, angle=-90}}
\caption{A conducting boundary (bold line). We indicate the
  integration surface $S$ normal to it
 and the 'pill box' type volume $V$ encompassing both sides.
Our equations become exact in the limit where the height of the pill box
 and the width of the surfaces $S$   approach zero.}
\label{f:pill}
\end{figure}

The homogeneous Maxwell equations require that the tangential part of the
electric field $E_\pa$ and the normal component of the magnetic field,
$B_\perp$,  are continuous across the boundary. This is usually derived
by the following argument:
denoting the two sides of the capacitor plate by the
super-scripts $+$ and $-$ the homogeneous Maxwell equations imply for
some surface $S$ spanning from one side to the other of our boundary
or for a volume $V$ encompassing a little of both sides of the boundary
(see Fig.~\ref{f:pill})
\be
\begin{array}{lllll}
0 &= & \int_S(\nabla\wedge E)d\si = &\int_{\dd S}E_\pa ds = &
 L(E_\pa^+ -E_\pa^-) \quad\mbox { hence}\\
 {}[E_\pa ] &\equiv & E_\pa^+ - E_\pa^- =& 0 ~, & \quad \mbox{ and}\\
0 &= & \int_V(\nabla B)dv = & \int_{\dd V} Bd\si = &
S(B_\perp^+ - B_\perp^-)     \quad\mbox { hence}\\
{} [B_\perp ] &\equiv & B_\perp^+ - B_\perp^- =& 0~. \\
\end{array}
\ee
In the same manner, integrating the inhomogeneous Maxwell
equations implies for the normal component of $\ve E$ and the
tangential component of $\ve B$
\bea
[E_\perp ] \equiv  E_\perp^+ - E_\perp^- &= & 4\pi\rho ~,\\
 {} [B_\pa ] \equiv  B_\pa^+ - B_\pa^- &= & 4\pi j \wedge n~.
\eea

Similar junction conditions exist also for Einstein's equations.
(Lanczos, 1922, Darmois 1927, Misner \& Sharp 1964, Israel 1966, see
Refs~\cite{Is,La,Da,Mi}), the so 
called junction conditions. To obtain them, we have to split the
geometrical quantities into components parallel and transverse to a
given hyper--surface. This is often also done in 4d gravity (3+1 formalism
or ADM formalism), where one wants to study the time evolution of the
metric on a 3d spatial hypersurface.  Examples are numerical
relativity, or canonical quantization of gravity where the canonical
fields are the spatial metric components $q_{ij}$ and their
canonical momenta are given by the extrinsic curvature, $\pi_{ij} = K_{ij}$.

There one considers  spacelike hypersurfaces, \ie, hypersurfaces with
timelike normal $n$,  $g(n,n)<0$. In braneworlds
we have a timelike  hypersurface with spacelike normals, $g(n,n)> 0$.

Using a slicing of spacetime into 4d hyper surfaces, one can express
the 5d Riemann curvature in terms of the 4d one and the extrinsic
curvature. The equations are relatively simple if we write them in Gaussian
coordinates (Gauss-Codazzi-Mainardi formulas, see \eg~\cite{Mi})
\bea
^5R^\mu_{~\nu\alpha\beta} &=& R^\mu_{\nu\alpha\beta} +K_{\nu\alpha}K^\mu_\beta
       - K_{\nu\beta}K^{\mu}_\alpha  \qquad \mbox{(Gauss formula)}
       \label{eq:Gauss}\\ 
^5R^4_{~\mu\nu\alpha} &=& \nabla_\nu K_{\mu\alpha} - \nabla_\alpha K_{\mu\nu}
        \qquad\qquad \quad\mbox{(Codazzi formula)}   \label{eq:Cod}\\
^5R^4_{~\mu 4\nu} &=& \dd_yK_{\mu\nu} + K_{\mu\beta} K^\beta_{\nu}
         \qquad\qquad \qquad \mbox{(Mainardi formula)} \label{eq:Main}\\
 ^5R  &=& {^5R^{\mu\nu}}_{\mu\nu} + 2~{{^5R}^{4\mu}}_{4\mu} =
   R + 2\dd_yK -K^2 -K_{\alpha\beta}K^{\alpha\beta}~.
\eea
For the last equation we have used $\dd_yg^{\mu\nu}=2K^{\mu\nu}$ so that
$$
g^{\mu\nu}\dd_yK_{\mu\nu} =\dd_y(g^{\mu\nu}K_{\mu\nu}) -
(\dd_yg^{\mu\nu})K_{\mu\nu} = \dd_yK - 2K^{\mu\nu}K_{\mu\nu}~, \qquad
 K =K^\mu_\mu~.
$$
From these we can determine the 5-dimensional Einstein tensor,
\bea
^5G^4_4 &=& \frac{1}{2}\left[- R + K^2 - K^{\mu\alpha} K_{\alpha\mu}\right]\\
^5G^4_\mu &=& \nabla_\mu K - \nabla_\nu K_\mu^\nu \\
^5G^\mu_\nu &=& G^\mu_\nu  + 2K^\mu_\alpha K^\alpha_\nu - K K^\mu_\nu
  + \dd_y K^\mu_\nu- \de^\mu_\nu\dd_yK +
\frac{1}{2} g_{\mu\nu}(K^2 +  K_{\alpha\beta}K^{\alpha\beta})~.
\eea

The derivatives wrt $y$ indicate that in order to determine the
5-dimensional Einstein tensor, it is not sufficient to know the
 second fundamental form on the brane itself, but we also have to
know it on both sides of the brane.
\vspace{0.6cm}

We now derive equations on the brane from the bulk Einstein equation,
 $G_{AB} = \ka_5 T_{AB}$ which we assume to be valid as a low energy
consequence from string theory. To identify the energy momentum
tensor on the brane which contains a delta-function in $y$-direction,
we define
$$
 S^A_B = \lim_{\epsilon \rightarrow 0}
\int_{y_b-\epsilon}^{y_b+\epsilon}T^A_Bdy~,
$$
so that
$$
\lim_{\epsilon \rightarrow 0}
\int_{y_b-\epsilon}^{y_b+\epsilon} {^5G}^A_Bdy = \kappa_5 S^A_B~.
$$
The 4d metric is continuous and also $K_{\mu\nu}$ has no
delta-function in $y$, but possibly a jump across the brane. This means that
${^5G}^4_4$ and ${^5G}^4_\mu$ have
 no delta-function. Only ${^5G}_{\mu\nu}$ may have one stemming from the term
$\dd_yK_{\mu\nu} -g_{\mu\nu} \dd_yK $ if $K_{\mu\nu}$ has a jump.
Hence as consequence from the junction conditions, we obtain the
following relations for $S_{AB}$
\bea
0 &=& S^4_4 \\
0 &=& S^4_\mu  \qquad \mbox{  and} \\
\kappa_5 S^\nu_\mu &=& \left[K^\nu_\mu\right]
-\delta^\nu_\mu\left[K\right]  \qquad  \mbox{ or } \\
  \left[K^\nu_\mu\right] &=& \kappa_5(S^\nu_\mu
  -\frac{1}{3}\delta^\nu_\mu  S)  \qquad \mbox{  where } \quad S
  =S^\mu_\mu = S^A_A~.  \label{e:junk}
\eea

\subsection{$Z_2$ symmetry}

In addition to Gaussian normal coordinates (which one can always
choose, at least locally)
we now assume $Z_2$ symmetry: the two sides of the brane are mirror
images. For the metric components this implies
 \bea
   g_{\mu\nu}(x^\lambda,y_b + y) &=& g_{\mu\nu}(x^\lambda,y_b - y)   \\
    g_{\mu 4}(x^\lambda,y_b + y) &=& - g_{\mu 4}(x^\lambda,y_b - y) \\
    g_{44}(x^\lambda,y_b + y) &=& g_{44}(x^\lambda,y_b - y)
\eea
The same symmetry is required for $T_{AB}$ and any other bulk tensor field.

Under this condition we have $K_+ = -K_-$   so that the junction
conditions (\ref{e:junk}) reduce to
\be
2K_\nu^\mu = \ka_5(S_\nu^\mu  -\frac{1}{3}\delta^\mu_\nu S) ~.
\ee
If the braneworld satisfies $Z_2$ symmetry, the brane energy momentum
tensor determines the second fundamental form.

However, we now show that even with  $Z_2$ symmetry, knowing the brane
energy momentum tensor is not
sufficient to determine the brane Einstein tensor. To demonstrate this
we first rewrite the 4d Einstein tensor, for a general coordinate system
in terms of the 5d Riemann tensor and the extrinsic curvature:
\bea
G_{\mu\nu} &=& ^5G_{AB}q_\mu^Aq_\nu^B + ^5R_{AB} n^An^Bg_{\mu\nu} -
K_\mu^\alpha K_{\alpha\nu} +  K K_{\mu\nu}  \nonumber \\     &&
+ \frac{1}{2} g_{\mu\nu}(K^2- K_{\alpha\beta}K^{\alpha\beta})  -
\tilde E_{\mu\nu} ~.   \label{e:Shiru}
\eea
Here $ ^5R_{AB} n^An^B$ corresponds to $^5R_{44}= {^5R}^{\mu}_{4\mu4}$ in
Gaussian coordinates and
$ \tilde E_{\mu\nu} \equiv ~{^5R}_{ABCD}n^An^Cq_\nu^Dq_\mu^B$ corresponds
to $^5R_{4\mu 4\nu}$. Eq.~(\ref{e:Shiru}) in Gaussian coordinates is a
simple consequence of the expressions
(\ref{eq:Gauss},\ref{eq:Cod},\ref{eq:Main}) for the 
Riemann tensor. In a general coordinate system it can be found in
Ref.~\cite{Shiru,RoyRev} (careful, the sign for $K_{\mu\nu}$ is different
there). In 5 dimensions the Weyl tensor is given by
\be \label{Weyl}
R_{ABCD} = \frac{2}{3}(g_{A[C}R_{D]B} - g_{B[C}R_{D]A}) - \frac{1}{6}
    g_{A[C}g_{D]B}R + C_{ABCD} ~.
\ee
Here $[A B]$ indicates anti-symmetrization in the indices $A$ and $B$, and
$C_{ABCD}$ is the 5-dimensional Weyl tensor defined by
Eq.~(\ref{Weyl}). It is easy to verify that $C_{ABCD}$ is traceless
and obeys the same symmetries as the Riemann tensor, $R_{ABCD}$.

Inserting the 5-dimensional Einstein Eq.~(\ref{eq:Ein}) for $^5G_{AB}$
and Eq.~(\ref{Weyl}) in the expression containing the Ricci tensor
$R_{AB}$, as well as in $\tilde E_{\mu\nu}$, we
obtain the 4d brane gravity equation
\bea \label{e:brgr}
G_{\mu\nu} &=& \frac{2}{3}\ka_5 \left[T_{AB}q_\mu^Aq_\nu^B +
  g_{\mu\nu} (T_{AB} n^An^B - \frac{1}{4}T)\right] - K_\mu^\alpha
  K_{\alpha\nu} + K K_{\mu\nu}
\nonumber \\   \label{eq:Ein4}   &&
 +\frac{1}{2}g_{\mu\nu}\left( K_{\alpha\beta}K^{\alpha\beta}-
 K^2\right)  - E_{\mu\nu}
\eea
where
\be \label{e:Weylel}
 E_{\mu\nu} = C_{ABCD}n^An^Cq_\nu^Dq_\mu^B
\ee
is the 'projection' of the Weyl tensor along the brane normal.
The Codazzi equation~(\ref{eq:Cod}) gives in addition
\be
\ka_5 T_{AB}n^Aq^B_\mu = \nabla_\mu K - \nabla_\nu K_\mu^\nu ~.
\ee

Because of the last term in Eq.~(\ref{eq:Ein4}), it is not sufficient to
know the bulk energy momentum tensor and initial conditions for $n$, $g_{AB}$
and $K_{AB}$ to solve the gravitational equations on the brane. In
addition we need to know $E_{\mu\nu}$, components of the bulk
Weyl tensor on the brane. The Weyl tensor, which is the part of the curvature
which can be non-vanishing even if the energy momentum tensor vanishes,
contains information on bulk gravity waves.
Bulk gravity waves can flow onto, respectively be emitted from the
brane and  thereby affect its evolution.
This information is encoded only in the full bulk initial conditions.
Therefore, to determine the evolution of the brane matter and geometry,
in principle we have to solve the full bulk equations!
Only in situations with very special symmetries this can be
avoided. However, as soon as we want to perturb such symmetric
solutions we have to take into account all the bulk modes, and we do
expect the solutions to differ significantly from the results of
4-dimensional perturbation theory. 

\subsection{Brane gravity with an empty bulk}
We now exemplify the effect of the bulk Weyl tensor in the case of an
empty bulk.
This will be the situation which we study for the rest of these lectures.
We assume that the bulk is empty up to a simple cosmological constant
$\Lambda_5$.
\be
    T_{AB} = -\frac{\La_5}{\ka_5} g_{AB}  +q_A{}^\mu q_B{}^\nu
                S_{\mu\nu}\de(y-y_b)~.
\ee
where  $S_{\mu\nu}$ is the energy momentum tensor on the brane. It consists
of a brane tension $\la$ and the matter energy momentum tensor $\tau_{\mu\nu}$.
\be
S_{\mu\nu}  = \la g_{\mu\nu} +  \tau_{\mu\nu}~.
\ee

The junction conditions read
\bea
\left[g_{\mu\nu}\right] &=& 0  \qquad \mbox{ first junction
  condition,}\\
   \left[K_{\mu\nu}\right] &=&
-\ka_5\left(S_{\mu\nu}  -\frac{1}{3}g_{\mu\nu}S \right) \qquad \mbox{
  second junction condition.} 
\eea
 $Z_2$ symmetry requires that
\be\label{e:junc}
    K^+_{\mu\nu} = - K^-_{\mu\nu} = -\frac{\ka_5}{2}\left(S_{\mu\nu}
    -\frac{1}{3}g_{\mu\nu}S \right)~.
\ee

Inserting our ansatz for $T_{AB}$ in the brane gravity
equation~(\ref{e:brgr}) and using the second junction
condition to eliminate the second fundamental form, we obtain
\be\label{e:Ein4}
G_{\mu\nu}  =  -\La_4 g_{\mu\nu}  + \ka_4 \tau_{\mu\nu}  +
  \ka_5^2\si_{\mu\nu}  - E_{\mu\nu}~,
\ee
with
\bea
\La_4 &=& \frac{1}{2}(\La_5 + \frac{\ka_5^2}{6}\la^2)~, \\
 \ka_4 &=& \ka_5^2\la/6 = 2/ M_4^2 \label{e:k45}\\
\si_{\mu\nu}   &=& - \frac{1}{4}\tau_{\mu\alpha}\tau^\alpha_\nu + \frac{1}{12}
    \tau\tau_{\mu\nu}  + \frac{1}{8}g_{\mu\nu}\tau_{\alpha\beta}\tau^{\alpha\beta}
  - \frac{1}{24}g_{\mu\nu}\tau^2~,
\eea
and, as before $E_{\mu\nu}$  denotes the projected 5d Weyl tensor,
evaluated on either side of the brane (but not exactly on the brane
where it may be ill-defined). The quantities $\ka_4$ and
$M_4$ denote the 4--dimensional gravitational coupling constant and Planck mass
respectively and $\tau = \tau_\mu^\mu$ is the trace of the matter
energy momentum tensor. The relation between the 4- and
5-dimensional Planck mass in the braneworld approach is now obtained
as follows: using $\ka_5 \sim M_5^{-3}$ and $\ka_4 \sim M_4^{-2}$,
Eq.~(\ref{e:k45}) shows that $M_4^2 \simeq M_5^6/\la$. Using that the
4-dimensional cosmological constant is small, $\La_4\ll |\La_5|$, we
have $\la \simeq \sqrt{-6\La_5}/\ka_5 \simeq  \sqrt{-6\La_5}M_5^3$, so that
$M_4^2 \simeq M_5^3\sqrt{-\La_5} = M_5^3L$ with $L \simeq \sqrt{-\La_5}$.

In the limit  $\la\tau_{\mu\nu} \gg \tau_{\mu\alpha}\tau^\alpha_\nu $
we recover the 4--dimensional Einstein equation if $E_{\mu\nu}$  is
negligible. The existence of this limit depends crucially on the
existence of a 4--dimensional brane tension. In order for the 4d
gravitational coupling constant to be positive, the brane tension
must be positive, $\la>0$. At high energy densities (in the early universe) the
quadratic term $\si_{\mu\nu}$  can become dominant and modify the
dynamics (the expansion law of the universe). In general,
there is an additional part, $E_{\mu\nu}$ , carrying information from the bulk
geometry and evolution, which can affect the brane evolution in a crucial way.

\subsection{Energy momentum conservation}
The Codazzi equation~(\ref{eq:Cod})  together with $Z_2$ symmetry implies
\be
          \nabla_\mu K -\nabla_{\nu} K_\mu^\nu   = \ka_5 T^4_\mu  = 0~.
\ee
With the Gauss equation~(\ref{eq:Gauss})  and $Z_2$ symmetry this ensures
energy and momentum conservation on the brane,
\be
        \nabla_{\nu} \tau_\mu^\nu = 0.
\ee
From the 4-dimensional contracted Bianchi identities we obtain in addition
\be \label{eq:Esi}
\nabla_\nu E_\mu^\nu  = \ka_5^2\nabla_\nu\si_\mu^\nu ~.
\ee
Hence, the longitudinal part of $E_{\mu\nu}$  is fully determined by the
matter content of the  brane, while the transverse traceless part is
not specified:
\be
  E_{\mu\nu}   = E_{\mu\nu}^{(TT)}  + E_{\mu\nu}^{(L)}
\ee
where    $\nabla_\nu E_\mu^{\nu~(TT)}= 0 $ and
\[
E_{\mu\nu}^{(L)}   = \frac{1}{2}(\nabla_\mu\theta_\nu  +
\nabla_\nu\theta_\mu ) \quad \mbox{  with } \quad \nabla_\mu\theta^\mu = 0~.
\]
Inserting this in Eq.~(\ref{eq:Esi}), we obtain
\be
\nabla^2\theta_\mu  = \frac{\ka_5^2}{2}\left[-\tau_{\alpha\beta}(\nabla_\mu
   \tau^{\alpha\beta} + \nabla^\alpha \tau_\mu^\beta) +\frac{1}{3}
(\nabla_\alpha \tau)(\tau_\mu^\alpha - q_\mu^\alpha \tau)\right]~.
\ee
For given initial conditions, this equation has always a unique
solution $\theta_\mu$
on the brane which determines $E_{\mu\nu}^{(L)}$. However, the transverse part,
$E_{\mu\nu}^{(TT)}$ is not determined by the brane energy momentum tensor; it
comes from bulk gravity waves. Only if $E_{\mu\nu}^{(TT)} = 0$ does the brane
energy momentum tensor determine the brane Einstein tensor. As we shall see,
already in quite simple situations this is not the case.

\section{The Randall Sundrum model}  \label{sec:RS}
 We now consider an Anti-de
Sitter (AdS) bulk, $\La_5 < 0$ and would like to obtain Minkowski space on
the brane. Since AdS is conformally flat, $E_{\mu\nu} = 0$. A Minkowski brane
can be achieved by setting $\tau_{\mu\nu} = 0$. If in addition the
brane tension is related to the 5--dimensional coupling constant and
the cosmological constant by
\be \label{eq:RSfin}
\la^2\ka_5^2/6 = - \La_5~,
\ee
Eq.~(\ref{e:Ein4}) implies $G_{\mu\nu} =
0$.  Eq.~(\ref{eq:RSfin}) is it the Randall--Sundrum (RS) fine tuning
condition~\cite{RS1,RS2}. Small deviations
from the RS condition lead to an exponentially expanding/contracting brane.
The 4-dimensional gravitational constant becomes
\be \label{eq:RSk4}
   \ka_4 = \frac{\la\ka_5^2}{6} = - \frac{\La_5}{\la}  > 0~,
 \quad \mbox{ or, equivalently }\quad    \la = - \frac{\La_5}{\ka_4}~,
 \quad \mbox{ and }\quad \ka_4 = \ka_5\sqrt{\frac{-\La_5}{6}} ~.
\ee

\subsection{The metric}
The following coordinates for (a part of) Anti-de Sitter will be useful
for us:
\bea
ds^2 &=& e^{-2|z|/\ell}\eta_{\mu\nu}dx^\mu dx^\nu + dz^2 \quad
\mbox{ Gaussian coordinates}  \label{eqRS:Gauss}\\
ds^2 &=& \left(\frac{\ell}{y}\right)^2
     \left(\eta_{\mu\nu}dx^\mu dx^\nu + dy^2\right)
 \qquad  |y| > \ell \quad \mbox{ conformal coordinates.} \label{eqRS:conf}
\eea
Einstein's equations,  $G_{AB} = -\La_5 g_{AB}$,  give
$\Lambda_5 = -\frac{6}{\ell^2}$.
The RS fine tuning requires $\la = \sqrt{- 6\La_5/\ka_5^2}$.

In their first model~\cite{RS1} (RS1 model) Randall and Sundrum
propose two branes, the first positioned at $z=0$ is called the hidden
brane, and the second positioned at $z=k\ell$ is called the visible brane
and represents our Universe. The gravitational force on the visible
brane is suppressed by the factor $\exp(-2k)$ w.r.t. the hidden brane,
leading to an enhancement by a factor  $\exp(k)$ of the apparent Planck mass.
However, also this two brane model contains a gravi-scalar (also
called radion) which has to
obtain a mass by some non-gravitational mechanism.

\begin{figure}[ht]
\centerline{\epsfig{figure=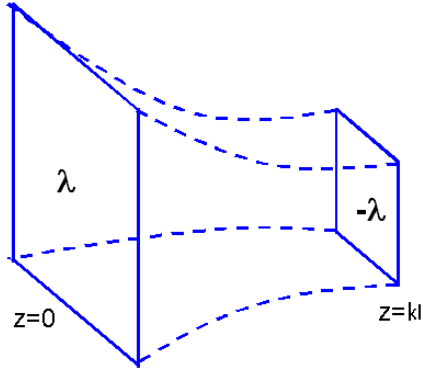, width=6.5cm}}
\caption{The RS1 model with two branes. The visible brane (our Universe) at $z=k\ell$
and the hidden brane at $z=0$.}
\label{f:RS2branes}
\end{figure}

\begin{figure}[ht]
\centerline{\epsfig{figure=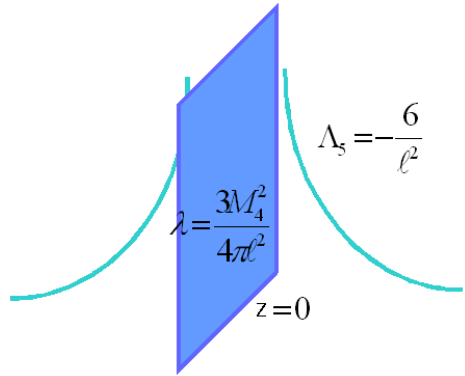, width=6.5cm}}
\caption{The RS2 model with only one brane at $z=0$.}
\label{f:RS1brane}
\end{figure}

This problem is resolved in the second model~\cite{RS2} (RS2 model). There, our
universe is located on the brane at $z=0$ corresponding to $y=\ell$ and
no second brane is present. The apparent 4-dimensional Planck mass as
measured on the brane on scales much larger than $\ell$ is then again given by
\bea\label{eqRS:M4}
\lambda &=& -\frac{\ka_4}{\La_5} = \frac{3M_4^2}{\ell^2}~. \qquad
\mbox{ Furthermore,}\\
\kappa_4^2 &=& -\frac{\Lambda}{6}\kappa_5^2 = \ell^{-2}\kappa_5^2 \qquad
\mbox{ so that}\qquad
M_4^2 = M_5^3\ell~.
\eea
Hence in the RS2 model, the AdS curvature scale $\ell$ enters
in the same way as $L$ for cylindric Kaluza-Klein models.
Since the scale $\ell$ is limited by present day micro gravity
experiments which have not detected any deviation from Newton's
law~\cite{micGra}, we have
$$
\ell < 0.1~mm \quad \mbox{ implying } \qquad \lambda > (1~TeV)^4, ~ \mbox{ hence }\quad
M_5 = (M_4^2 /\ell)^{1/3} > 10^5 ~TeV~.
$$

\subsection{Gravity waves in the RS2 model}
In order to see that the radion mode is absent in the non-compact RS2 model,
we consider perturbations to the AdS metric. It is easy to show that one can
always choose a gauge (local coordinate system) so that the perturbed metric
is of the form
\begin{eqnarray}
ds^2 &=& \left(\frac{\ell}{y}\right)^2\left[-(1+2\Psi)dt^2
-4\Sigma_idtdx^i +((1-2\Phi)\delta_{ij}
   + 2H_{ij})dx^idx^j  +4\Xi_idydx^i \right.\\
 && \left. \qquad \quad -4{\cal B}dtdy +(1+2{\cal C})dy^2 \right]
\end{eqnarray}

Here $H_{ij}$ and $\Sigma_i$, $\Xi_i$ are transverse (\ie divergence
free) and $H_{ij}$ is traceless.
In other words, $\dd_i\Sigma_i =\dd_i\Xi_i =\dd_iH_{ij} =H_i^i =0$.
As we shall see there are 5 homogeneous modes in these
10 physical perturbation variables corresponding to the
5 gravity wave modes in 5 dimensions.
We consider the homogeneous 5d wave equation in the bulk.

Since the 3-brane is homogeneous, scalar-, vector- and tensor
degrees of freedom decouple and we can consider them in
turn (for more details see Section~\ref{sec:cos4}).

\subsubsection{Tensor perturbations}
We first consider only the tensor $H_{ij}$, so that the perturbed metric is
given by
\be
ds^2 = \left(\frac{\ell}{y}\right)^2\left(-dt^2 +(\delta_{ij}
   + 2H_{ij})dx^idx^j  +dy^2 \right)~.
\ee
We Fourier transform $H_{ij}$ in the 3-dimensional $\bx$ coordinates
and consider one mode with fixed wave vector $\bk$, so that
$H_{ij}(t,y,\bx) = H_{ij}(t,y)\exp(i\bk\cdot\bx)$. Since spacetime is
isotropic and homogeneous in $\bx$, different $\bk$--modes do not couple.
The bulk Einstein equations, $\de G_{{AB}} = -\Lambda \de g_{AB}$,
for the Fourier mode $k$ then give
\be
\left(\dd_t^2 +k^2 -\dd_y^2 + \frac{3}{y}\dd_y \right)
H_{ij} = 0 ~.
\ee
The general solution to this equation is of the form
\bea
H_{ij} = h_m e_{ij} \quad \mbox{ with } 
h_m &=& e^{i\omega t}(my)^2\left[AJ_2(my) +BY_2(my)\right]~.
\eea
Here $e_{ij}$ is the polarization tensor, $k^ie_{ij} = e^i_{~j}=0$ and
$\om^2 = m^2 + k^2$. The separation constant $m^2$ is arbitrary and can,
in principle also be negative. $J_2$ and $Y_2$ are the Bessel functions
of order 2. They are oscillating and decaying. Bessel functions
represent ``$\de$--function 
normalizable'' perturbations like harmonic waves in flat space, in the
sense that~\cite{Csaki:2000fc,Bozza:2001xt}
\begin{equation}
 \int_0^\infty h_mh_{m'} \frac{dy}{m^2y^3} = m\de(m-m') ~.
 \label{e:normalisation}
\end{equation}
These are just the ordinary gravity modes of $4$-dimensional mass $m$
without a mass gap which are discussed in the original RS
paper~\cite{RS2}.  To find the correct weight $1/y^3$, we use that $h_m$
satisfies
\be
\left(\Box{ }_4 +\dd_y^2 -\frac{3}{y}\dd_y\right)h_m =0~,
\ee
and thus $\check{h}= h_m/y^{3/2}$ satisfies the equation of motion of a scalar
field in a flat 5-dimensional spacetime (with $y$--dependent mass
term),
\be
\left(\Box{ }_4 +\dd_y^2 -\frac{15}{4y^2}\right)\check h =0~.
\ee
 This mode has  to be
normalizable w.r.t the Minkowski metric (no additional weight).

\begin{figure}[ht]
\centerline{\epsfig{figure=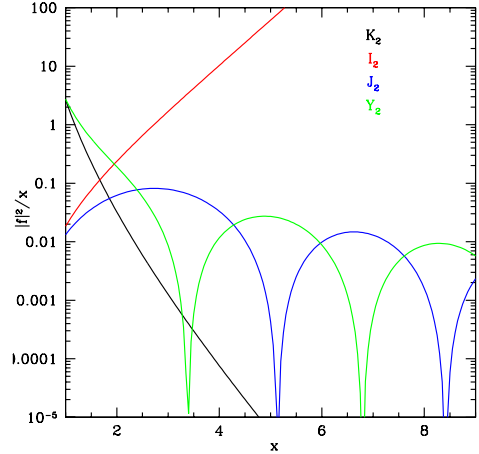, width=6.5cm}}
 \caption{The Bessel functions of order 2.}
\label{f:bess}
\end{figure}

As we have mentioned above, $m^2$ is arbitrary and can also be chosen negative.
However, if $m^2<0$ and therefore $m$ is imaginary, it is more
useful to decompose the two independent solutions in the form
\be
h_m = e^{i\omega t}(|m|y)^2\left[CK_2(|m|y) +DI_2(|m|y)\right]~,
\ee
where $K_2$ and $I_2$ are the modified Bessel functions of order 2.
Considering the behavior of the Bessel
functions, one sees that $I_2$ grows exponentially (see Fig.~\ref{f:bess})
and  is clearly not normalizable (\ie not square integrable with some weight
which is a power law in $y$). Therefore, this mode is unphysical and
we have to set $D=0$. It is important to note that $\omega^2=k^2+m^2$
can become negative in this case leading to $\omega =\pm i|\om|$. In
other words, negative mass solutions become exponentially growing
'tachyonic' instabilities! It is still unclear whether these tachyonic
modes are relevant for cosmological braneworlds.

In the limit $m\rightarrow 0$ only the $Y_2$--mode survives and we obtain
$ h_{(m=0)} =C\exp(\omega t)$ independent of the coordinate
$y$. This zero-mode is normalizable with respect to the measure
$dy/y^3$.

The general solution for a tensor perturbation is  of the form
\be
h=h_0 + \int_{-\infty}^{\infty}h_mdm^2~.
\ee
At the brane position, $y=\ybr=\ell$, the perturbations must satisfy
the junction condition~(\ref{e:junc}). These represent boundary
conditions for the perturbations $H_{ij}$ in the bulk. On the right
hand side of Eq.~(\ref{e:junc}), we can in principle have an arbitrary
perturbation of the matter energy momentum tensor. However, the only
non-vanishing term of the 
tensor contribution to $\tau_{\mu\nu}$ is the traceless part of
$\tau_{ij}$, i.e. the anisotropic stress on the brane, $\Pi^{(T)}_{ij}$.
A short computation shows
\bea
  \left. \de K_{ij} \right|_{\ybr}
   & =& \left.\left(\frac{2}{\ell}H_{ij}-\dd_y H_{ij}\right)\right|_{\ybr}~,
    \quad\mbox{hence}  \nonumber \\  \label{eq:2ndten}
 - \left. 2(\dd_y H_{ij}) \right|_{\ybr}
   & = & \kappafive \Pi^{(T)}_{ij}~,
\eea
where $\Pi^{(T)}$ are tensor--type anisotropic stresses on the brane.

Let us first consider the homogeneous case $\Pi^{(T)}\equiv0$.  For
$m^2>0$, the solutions are of the form
\begin{equation}
 h = \exp(\pm i\om t)(my)^2\left[ AJ_2(my) +BY_2(my)\right]  ~.
\end{equation}
The junction condition~(\ref{eq:2ndten}) then requires
\begin{equation}
 B = -A\frac{J_1(m\ell)}{Y_1(m\ell)} \simeq \frac{\pi}{4}(m\ell)^2A ~,
\end{equation}
where the last expression is a good approximation for $m\ell\ll 1$. This
is precisely the result of Randall and
Sundrum~\cite{RS2}. It is not modified even if we allow
for the negative mass modes, $-m^2 >0$, because a physical solution
has to be of the form
\begin{equation}\label{eq:tachten}
  h = C\exp(\pm t\sqrt{-m^2-k^2})(|m|y)^2K_2(|m|y)~, \quad \mbox{ with }
  \quad  \dd_yh = |m|C\exp(\pm t\sqrt{-m^2-k^2})(|m|y)^2K_1(|m|y)~,
\end{equation}
and since $K_1$ has no zero, the junction condition (\ref{eq:2ndten})
requires $C=0$.

But in a realistic brane universe, $\Pi^{(T)}$ is not exactly zero. In
cosmology, it is  typically just a factor 10 smaller than other
perturbations of the energy momentum tensor on the brane. We therefore
cannot require $C\equiv 0$. However, as long as $\Pi^{(T)}$ remains
small, we do not expect the unstable modes to be present, so that
$C(k,m)=0$ for $k^2<-m^2$.  Within the framework of first order
perturbation theory, the $\Pi^{(T)}$ modes satisfy a Minkowski
equation of motion and therefore they do not grow exponentially. Hence
in this case, the $K$--mode  can only be excited for
$\om^2=k^2+m^2>0$, \ie ~$k^2>-m^2$. However, it is not clear
whether this remains true to second order, where the evolution of $H$
feeds back in the equation of motion for  $\Pi^{(T)}$.
Actually, within a toy model, it has been shown that the full, non-linear
evolution can be exponentially unstable even if the linear equations
do not excite the unstable mode~\cite{CD4}.

\subsubsection{Vector perturbations}
We now consider vector perturbations only, so that, in generalized
longitudinal gauge, the metric takes the form

\begin{eqnarray}
ds^2 &=& \left(\frac{\ell}{y}\right)^2\left(-dt^2
-4\Sigma_idtdx^i +\delta_{ij}dx^idx^j  +4\Xi_idydx^i
+ dy^2 \right)
\end{eqnarray}

The bulk Einstein equations for a mode $\bk$ of the vector
perturbations $\ve\Si$ and $\ve\Xi$ are
\begin{align}
 \left(\dd_y^2-\frac{3}{y}\dd_y\right)\ve\Si
 &= \left(\dd_t^2+k^2\right)\ve\Si~, \label{eq:bulkvecSi}\\
 \left(\dd_y^2-\frac{3}{y}\dd_y+\frac{3}{y^2}\right)\ve\Xi
 &=\left(\dd_t^2+k^2\right)\ve\Xi~,\label{eq:bulkvecXi}\\
 \left(\dd_y -\frac{3}{y}\right)\ve\Xi 
 & = - \dd_t\ve{\Si}~,\label{eq:bulkvecconst} 
\end{align}
where $\ve\Si$ and $\ve\Xi$ are transverse vectors,
$(\bk\cdot\ve\Si) =(\bk\cdot\ve\Xi) =0$.  The
constraint equation~(\ref{eq:bulkvecconst}) fixes the relative
amplitudes of $\Si$ and $\Xi$, showing that there is only one
independent vector perturbation in the bulk (the ``gravi-photon'').
One can check that these equations are consistent, e.g. with the master
function approach of Ref.~\cite{Mukohyama:2000ui}.

As in the tensor case, the solutions are Bessel functions of order
two (and one). Considering just one component $\Si=\Si_i$ one obtains
the expected oscillatory modes for positive mass-squared, $m^2>0$,
\begin{align}
 \Si &= \exp(\pm i\om t)(my)^2\left[A J_2(my) + BY_2(my) \right]~, \\
 \Xi &= \frac{\pm i\om}{m}  \exp(\pm i\om t)(my)^2\left[A
 J_1(my) +  BY_1(my) \right]~,
\end{align}
where $\om =\sqrt{m^2+k^2}$. These solutions have been found in
Ref.~\cite{Bridgman:2000ih}. For a negative mass-square,
$m^2< 0$, we obtain again tachyonic solutions. Like in the tensor
case, the solution containing the modified Bessel function $I_\nu$
cannot be accepted as it is exponentially growing and thus represents
a non-normalizable mode. However, the $K_\nu$-solution is
exponentially decaying and  perfectly acceptable. For
tachyonic vector perturbations with $\om^2 = m^2+k^2<0$ we have
\begin{align}
 \Si &= C\exp(\pm |\om| t)(|m|y)^2K_2(|m|y)~,  \\
 \Xi &= \frac{\pm |\om|}{|m|} C\exp(\pm |\om| t)(|m|y)^2K_1(|m|y)~.
\end{align}
For large enough scales, $-m^2>k^2$, these solutions again grow
exponentially.

 The boundary conditions at the brane relate these
perturbations to the brane energy momentum tensor. For the
energy momentum tensor on the brane, the vector degrees of freedom
are defined according to
\begin{equation}
 \left(S_{\mu\nu} \right)
 = \begin{pmatrix}0 & V_j \cr V_i & \Pi^{(V)}_{ij}\end{pmatrix}
   -\tension \left(q_{\mu\nu}\right) ~,
 \label{e:T_vector}
\end{equation}
where $V_i$ and $\Pi^{(V)}_i$ are divergence-free vector fields
and $\Pi^{(V)}_{ij}\equiv  [\dd_j \Pi^{(V)}_i+\dd_i
\Pi^{(V)}_j]$. The first junction condition simply requires that
$\Si$ be continuous at the brane, which it is since the (modified)
Bessel functions of even index are even functions. The second
junction condition results in (for a detailed derivation,
see~\cite{Ringeval:2003na})
\begin{align}
 \dd_t\Xi+\dd_y \Sigma &= \kappafive V~, \label{eq:jv1}\\
 \Xi &= \kappafive \Pi^{(V)}~,\label{eq:jv2} \\
 \dd_t V &=-k^2 \Pi^{(V)}~.
\end{align}
The last equation follows from (\ref{eq:jv1}) and (\ref{eq:jv2}) and
the bulk equations
(\ref{eq:bulkvecSi})--(\ref{eq:bulkvecconst}). It represents momentum
conservation on the brane, which is guaranteed as long as we have
vanishing energy flux off the brane and $Z_2$--symmetry.

Like for tensor perturbations, we consider homogeneous
solutions, setting $\Pi^{(V)}\equiv V \equiv 0$. This requires
$\Xi(|m|\ell)=0$, hence
\begin{align}
 B &= -A\frac{J_1(m\ell)}{Y_1(m\ell)} \quad\mbox{ for } m^2>0~,
 \label{e:B-vector-onbrane}\\
 C  &\equiv 0  \quad \mbox{ for } m^2<0~.
 \label{e:C-vector-onbrane}
\end{align}
Equation~(\ref{eq:jv1}) is then identically satisfied.

However, it seems more realistic to allow a small but non-vanishing
anisotropic stress contribution $\Pi^{(V)}$ and corresponding
vorticity $V$. In this case, again, we can have solutions with $C\neq
0$ which can grow exponentially in time; hence small initial data can
lead to an exponential instability like for tensor perturbations.

Using the normalization condition (\ref{e:normalisation}) for the
$m=0$ mode of the variable $\Xi\propto y$ (this is the one which
enters as dynamical variable in the perturbed action,
see~\cite{Csaki:1999jh}), one finds that $\int|\Xi|^2/y^3dy$ diverges
logarithmically. Contrary to the tensor case,
the vector zero-mode is not normalizable. Therefore, on the brane
there is only the ordinary massless spin--2 graviton, but there are a
continuous infinity of massive spin--2 and spin--1 particles (the
modes discussed here, with $m\neq 0$).

\subsubsection{Scalar perturbations}
We now discuss the most cumbersome, the scalar sector. Scalar--type
metric perturbations in the bulk are of the form
\begin{align}\label{eq:scal}
 d s^2
 &= \frac{\ell^2}{y^2}
     \left[-(1+2\Psi)dt^2  -4\BB dtdy \right.  \nonumber  \\
 &\quad \left.+(1-2\Phi)\delta_{ij}dx^i dx^j
    +(1+2\CC) dy^2 \right]~.
\end{align}

The bulk Einstein perturbation equations for the mode $\bk$ become,
after some manipulations and introducing the combination
$\Ga\equiv\Phi +\Psi $  ~~(see Ref.~\cite{CD4}),
\begin{align}
 \Phi - \Psi &=\CC~,   \label{eq:psi}\\
 \left(\dd_y^2 -\frac{3}{y}\dd_y\right)\Ga
 &= \left(\dd_t^2 +k^2\right)\Ga ~, \label{eq:ga}\\
 \left(\dd_y^2 -\frac{3}{y}\dd_y + \frac{4}{y^2}\right)\CC
 &= \left(\dd_t^2 +k^2\right)\CC  ~, \label{eq:c}\\
 \dd_y\Phi+ \left(\dd_y -\frac{3}{y}\right)\CC
 &= -\dd_t\BB ~,\label{eq:b}\\
 \frac{3}{y}\left(\dd_y-\frac{2}{y}\right)\CC
 &= 3\dd^2_t\Phi+k^2(\Phi +\CC)~,\label{eq:10}\\
 3\dd_t\left(\dd_y\Phi - \frac{1}{y}\CC\right)
 &= k^2\BB~, \label{eq:blast}\\
 \dd_t\left(2\Phi-\CC\right)
 &= \left(\dd_y -\frac{3}{y}\right)\BB~. \label{eq:last}
\end{align}

Clearly these equations are not all independent,
Eqs.~(\ref{eq:blast})--(\ref{eq:last}) are identically satisfied
if Eqs.~(\ref{eq:psi})--(\ref{eq:10}) are. The solutions are
obtained as for tensor and vector perturbations. For a positive mass-square,
$m^2 > 0$, we find ($\om =\sqrt{m^2+k^2}$)
\begin{align}
 \Ga &= \exp(\pm i\om t)(my)^2\left[A'J_2(my) + B'Y_2(my) \right]~, \\
 \CC &= \exp(\pm i\om t)(my)^2\left[AJ_0(my) + BY_0(my) \right] ~,\\
 \Phi &= \frac{1}{2}\exp(\pm i\om t)(my)^2
          \left[A'J_2(my)+ B'Y_2(my) \right.  \nonumber \\
      & \qquad \left. +AJ_0(my) +  BY_0(my) \right]~, \\
 \Psi &= \frac{1}{2}\exp(\pm i\om t)(my)^2\left[A'J_2(my)+ B'Y_2(my)
          \right. \nonumber \\
      & \qquad\left. - AJ_0(my) -  BY_0(my) \right]~, \\
 \BB &= \frac{\pm im^3y^2}{2\om} \exp(\pm i\om t)\times  \nonumber \\
 & \qquad \left[(A'-3A)J_1(my)+(B'-3B)Y_1(my) \right]~, \\
 \mbox{with }\quad A' &= 3A\frac{m^2}{m^2+2\om^2}~, \quad \mbox{ and } \quad
 B' = 3B\frac{m^2}{m^2+2\om^2}~.
\end{align}

 For a negative mass-square, $m^2 < 0$, we obtain ($\om =\sqrt{-m^2-k^2}$)
\begin{align}
 \Ga &= \exp(\pm \om t)(|m|y)^2C'K_2(|m|y)~, \\
 \CC &=  \exp(\pm \om t)(|m|y)^2CK_0(|m|y)~,\\
 \Phi &= \frac{1}{2}\exp(\pm\om t)(|m|y)^2 \times  \nonumber \\
 & \qquad \left[C'K_2(|m|y)+ CK_0(|m|y) \right] ~,\\
 \Psi &= \frac{1}{2}\exp(\pm \om t)(|m|y)^2\times  \nonumber \\
 & \qquad \left[C'K_2(|m|y) - CK_0(|m|y) \right]~, \\
 \BB &=\frac{\pm |m|^3y^2}{2\om} \exp(\pm \om t)[C'+3C]K_1(|m|y)~, \\
 \mbox{with }\quad C'&=-3C\frac{|m|^2}{|m|^2+2\om^2}~,
\end{align}
where we have already used that the $I$--mode is not normalizable and
therefore cannot contribute. Like for vector and tensor perturbations,
we find again  tachyonic solutions with $m^2<0$ which represent an
exponential instability for sufficiently small wave numbers $k$ (large
scales).

Determining the boundary conditions via the first and second
junction conditions now requires a bit more care. Since we have
already fully specified our coordinate system by the adopted
choice of perturbation variables, we must allow for brane bending.
We cannot fix the brane at $\ybr=\ell$, but we must allow for
$\ybr^{+}=\ell+\EE$ and $\ybr^{-}=-\ell-\EE$,
respectively. Fortunately, $\EE$ is a scalar quantity and brane
bending therefore does not affect vector and tensor perturbations. The
anti-symmetry $\ybr^{+}=-\ybr^{-}$ is an expression of
$Z_2$--symmetry. The introduction of the new perturbation
variable $\EE(x^\mu)$ describing brane bending enters the
exressions for the first and second fundamental forms. From
Eq.~(\ref{eq:firstfund}), 
we obtain $q_{\mu\nu}=g_{\mu\nu}$ to first order, which implies
that $\Phi$ and $\Psi$, hence $\CC$, have to be continuous. At the
brane position, the perturbed components of the extrinsic
curvature~(\ref{eq:Kmn}) are
\begin{align}
 \de K_{00}
   &= \frac{1}{\ell}\left[\Phi-3\Psi+ 2\frac{\EE}{\ell}\right]
      +\dd_y \Psi-2\dd_t \BB+\dd_t^2 \EE~,
      \label{e:K00-scalar} \\
 \de K_{0j}
   &= \dd_j\left(\dd_t \EE- \BB \right)~,
      \label{e:K0j-scalar} \\
 \de K_{ij}
   &= \left[\frac{1}{\ell}\left(\Psi -3\Phi -2\frac{\EE}{\ell}\right)
      +\dd_y \Phi\right]\delta_{ij}+\dd_i \dd_j \EE~.
      \label{e:Kij-scalar}
\end{align}
For the energy momentum tensor on the brane, we parameterize the
four degrees of freedom according to
\begin{equation}
 \left(S_{\mu\nu}\right)
 =  \begin{pmatrix}\rho & v_j \cr v_i & p\delta_{ij}
                   +\Pi^{(S)}_{ij} \end{pmatrix}
    -\tension  \left(q_{\mu\nu}\right) ~,
 \label{e:T_scalar}
\end{equation}
where $v_i \equiv \dd_i v$ and $\Pi^{(S)}_{ij} \equiv \left(\dd_i
\dd_j - \frac{1}{3}\Delta \delta_{ij}\right)\Pi^{(S)}$. With
Eqs.~(\ref{e:K00-scalar})--(\ref{e:Kij-scalar}), the second
junction condition reads
\begin{align}
 \frac{1}{\tension}\left(2\rho + 3p\right)
 &= \Phi\! -\! \Psi\! +\! \ell\dd_t\left(\dd_t\EE-2\BB\right)\!
     + \! L\dd_y \Psi~, \label{e:dK00} \\
 \frac{3}{\tension \ell}v
 &= \dd_t\EE-\BB~, \label{e:dK0j} \\
 \frac{3}{\tension \ell}\Pi^{(S)}
 &= \EE~, \label{e:dKij}\\
 \frac{1}{\tension}\left[\rho - \lap\Pi^{(S)}\right]
 &= \Psi-\Phi +\ell\dd_y\Phi ~. \label{e:dKii}
\end{align}
Combining the time derivative of Eq.~(\ref{e:dK0j}) with
Eqs.~(\ref{e:dK00}), (\ref{eq:b}) and (\ref{e:dKii}), we obtain
momentum conservation on the brane,
\begin{equation}
 \dd_t v = \frac{2}{3}\Delta \Pi^{(S)}+p~.
 \label{e-scalar-const-dv}
\end{equation}
Similar manipulations imply energy conservation on the brane,
\begin{equation}
 \dd_t\rho = \lap v~.
\end{equation}

Like for tensor and vector perturbations, we look for
solutions with vanishing brane matter. Setting $\Pi^{(S)}\equiv
\rho\equiv P\equiv v \equiv 0$ forbids brane bending, $\EE=0$.
Then Eq.~(\ref{e:dK0j}) implies $\BB(m\ell)=0$, thus
\begin{align} \label{e:BB'-scalar-onbrane}
 B'-3B &= -(A'-3A)\frac{J_1(m\ell)}{Y_1(m\ell)}
 \quad\mbox{ for }\quad  m^2>0~, \\
 C'+3C &= 0  \quad \mbox{ for }\quad  m^2<0~.
\end{align}
The other equations are all satisfied if we require separately
\begin{align}
 \frac{B}{A} &= \frac{B'}{A'}= -\frac{J_1(m\ell)}{Y_1(m\ell)}
 \quad\mbox{ for }\quad m^2>0~,
 \label{e:B-scalar-onbrane}\\
 C  &=  C' \equiv 0  \quad \mbox{ for }\quad  m^2<0~.
 \label{e:C-scalar-onbrane}
\end{align}
Since $B/A=B'/A'$, equations (\ref{e:BB'-scalar-onbrane}) are
(\ref{e:B-scalar-onbrane}) are equivalent.

As for vector perturbations, the $m=0$ scalar mode is not
normalizable. Like for tensor and vector perturbations, we have found
``scalar gravitons'' which appear on the brane as massive
particles. If the brane matter is unperturbed, only oscillating
$m^2>0$ solutions are possible. However, if we allow for non-vanishing
matter perturbations on the brane, we can have $C\neq 0$ and the
tachyonic modes $m^2<0$ can appear exactly like in the tensor and
vector sectors.

It is not surprising that the same instability appears in the scalar,
vector and tensor sectors, because all modes describe the same bulk
particle, the five-dimensional graviton.

\subsection{Green's function, correction to the Newtonian potential}
We want to determine the modification to Newton's law in the RS2 model.
Since the extra dimension is not compact and there are massive
(homogeneous) modes of all masses $m^2 > 0$, we expect a modification which
is not exponentially suppressed.

The 5d Green's function is defined by
$$ \nabla^2G(x,x') = \delta^5(x-x')~, $$
where $ \nabla^2$ is the 5d  d'Alembertian in AdS spacetime and we have
to glue together a $Z_2$--symmetric solution on both sides which satisfies
the homogeneous junction condition.
We can obtain the retarded Green's function in the standard way from
the homogeneous solutions of the equation (see \eg~\cite{CoHi}):
$$
G_R(x,x') = - \int\frac{d^4k}{(2\pi)^4}
    e^{ik_\mu(x^\mu-x^{\prime\mu})}\left[\frac{y^{-2}y^{\prime -2}\ell^{3}}{
   \ve{k}^2 -(\omega +i\epsilon)^2}  + \int_0^\infty dm
\frac{u_m(y)u_m(y')}{m^2+\ve{k}^2 -(\omega +i\epsilon)^2}\right]~.
$$
The first term comes from the $m=0$ solution and the functions $u_m$ are
the properly normalized massive modes,
$$ u_m(y) =\sqrt{\frac{m\ell}{2}}\frac{J_1(m\ell)Y_2(my) -Y_1(m\ell)J_2(my)}{
   \sqrt{J_1(m\ell)^2 + J_1(m\ell)^2}}~. $$

\begin{figure}[ht]
\centerline{\epsfig{figure=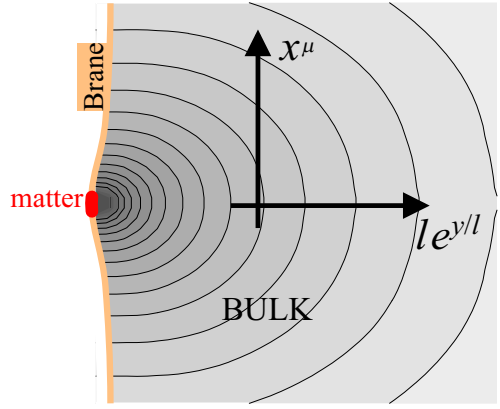, width=6.5cm}}
\caption{The equi-potential lines for the gravitational potential from a
point mass on the brane in RS2. Figure from~\cite{TG}. }
\label{f:Massbrane}
\end{figure}

The general retarded solution for a given energy momentum tensor
$\tau_{\mu\nu}$ on the brane  is now of the form
$$       h_{\mu\nu}(x) = -2\kappa_5\int d^4x' ~ G_R(x,x') S_{\mu\nu} (x')~.$$
For a stationary matter distribution  it is simpler to use the Green's
function of the spatial Laplacian which is related to $G_R$ via
integration over time
\bea
G(\bx,y,\bx',y') &=&\int_{-\infty}^{\infty}dt'G_R(x,x') =
- \int\frac{d^3k}{(2\pi)^3}
    e^{i\bk \cdot(\bx-\bx')}\left[\frac{y^{-2}y^{\prime -2}\ell^{3}}{
   \bk^2}  + \int_0^\infty dm
\frac{u_m(y)u_m(y')}{m^2+\bk^2 }\right]~,\\
 &=& \frac{-y^{-2}y^{\prime -2}\ell^{3}}{4\pi r} + \frac{1}{2\pi}
     \int_0^\infty dm {u_m(y)u_m(y')}\exp(-mr) ~.
\eea
On the brane, $y=y'=\ell$, the first term gives the usual $1/r$
behavior, $r=|\bx-\bx'|$.
Expanding the second term to lowest order in $\ell/r$ we obtain
$$
G(\bx,\ell,\bx',\ell) \simeq \frac{-1}{4\pi \ell r}\left[1 +
  \frac{\ell^2}{2r^2} + \cdots \right] ~.
$$
This determines  the Newtonian potential of a point mass on the brane
with mass $M$,
\be
\kappa_5 MG \simeq \frac{-\kappa_4M}{4\pi r}\left[1 +
  \frac{\ell^2}{2r^2} + \cdots \right]~.
\ee
Since the extra dimension is non-compact, the correction is not
exponentially suppressed but only as a power law.
Away from the wall, the potential at large separation is given by
\be
G(\bx,\ell,\bx',\ell+y) \simeq \frac{-\ell}{8\pi (\ell+y)^2}
  \frac{2r^2+ 3y^2}{(r^2+y^2)^{3/2}}
\ee
The equipotential lines are shown in Fig~\ref{f:Massbrane}. This
formulas have been derived in Ref.~\cite{TG} from which also
Fig.~\ref{f:Massbrane} is drawn.

\section{Cosmological perturbation theory in 4 dimensions}\label{sec:cos4}
Before studying brane cosmology and 5d effects on cosmological
perturbations, I present a brief introduction to 4d cosmological
perturbation theory and some aspects of 4d cosmology. Much more
details can be found \eg in~\cite{Rfund,Dod}. Some knowledge
of 4d cosmology  is however assumed (Friedmann equations etc. as they
can be found in the first chapter of standard textbooks on cosmology like
Refs.~\cite{peebles} or~\cite{Dod}).
Students who are familiar with this subject may skip this section.

\subsection{ Perturbation variables}
 \index{perturbation variables}
The observed Universe is not perfectly homogeneous and
isotropic. Matter is arranged in galaxies and clusters of galaxies and
there are large voids in the distribution of galaxies. Let us assume,
however, that these inhomogeneities lead only to small variations of
the geometry which we shall treat in first order perturbation
theory. For this we define the perturbed geometry by
\be
g_{\mu\nu} = \bar{g}_{\mu\nu}+ a^2 h_{\mu\nu}~,\qquad
\bar{g}_{\mu\nu}dx^\mu dx^\nu =a^2\left(-d\eta^2 +\ga_{ij}dx^idx^j\right) ~.
\ee
Here $\bar{g}_{\mu\nu}$ is the unperturbed Friedmann metric, $a(\eta)$
is the scale factor, $\eta$
denotes conformal time and $\ga_{ij}$ is the 3d metric for a space of
constant curvature $K$. The perturbations are assumed to
be small, $|h_{\mu\nu}|\ll 1$. The energy momentum tensor is given by
\be \begin{array}{llll}
T^\mu_\nu = \overline{T}^\mu_\nu + \theta^\mu_\nu,&
    \quad \overline{T}^0_0 = -\bar{\rho},&
    \quad \overline{T}^i_j = \bar{p} \de^i_j&
    \quad |\theta^\mu_\nu|/\bar{\rho} \ll 1 .
\end{array} \ee

The background energy density $\rho$ and pressure $p$ satisfy the
Friedmann equations,
\bea
\HH^2 \equiv \left(\frac{\dot a}{a}\right)^2 +K &=& \frac{8\pi
  G_4}{3}a^2\bar\rho +\frac{1}{4}a^2\La_4 \\
 \dot\rho = -3(\bar\rho+\bar{p})\HH~,
\eea
where an over-dot denotes the derivative w.r.t. conformal time $\eta$.

Without loss of generality we can choose the so-called longitudinal gauge
so that perturbations of the metric are of the form
\be
\left(h_{\mu\nu}\right) =  \begin{pmatrix}  -(1+2\Psi) &  B_i \cr
   B_i & (1-2\Phi)\ga_{ij} + H_{ij}
\end{pmatrix}~.
\ee
Here $B_i$ is an divergence free vector and $H_{ij}$ is a trace-free,
divergence free tensor field. The scalar quantities $\Psi$ and $\Phi$
are called the Bardeen potentials. In the Newtonian approximation they
are both equal and reduce to the Neewtonian potential.

We also decompose the perturbations into different 'Fourier modes',
$$
\Psi(\eta,\bx)  = Y_{\bk}(\bx)\Psi(\eta,\bk), \quad
\Phi(\eta,\bx)  = Y_{\bk}(\bx)\Phi(\eta,\bk), \quad
B^i(\eta,\bx)  = Y_{\bk}^{(V)i}(\bx)B(\eta,\bk), \quad
H_{ij}(\eta,\bx)  = Y_{\bk~ij}^{(T)}(\bx)H(\eta,\bk)~.
$$
In a Friedmann Universe with vanishing curvature, these are just
ordinary Fourier modes, while in the general case, the functions
$Y_\bk$ are eigenfunctions of the spatial Laplacian with eigenvalue
$-k^2$. The functions $Y_{\bk}^{(V)i}$ and $Y_{\bk~ij}^{(T)}$
correspondingly are vector- and tensor-type eigenfunctions of the
spatial Laplacian with vanishing divergence. For later use we also
define a scalar type vector and tensor as well as a vector-type tensor,
\bea
Y_{\bk~i}^{(S)} &=& -k^{-1}\nabla_i Y_{\bk}~, \qquad
Y_{\bk~ij}^{(S)} = k^{-2}(\nabla_i\nabla_j -\frac{1}{3} \de_{ij}\De)
Y_{\bk}~, \\ 
Y_{\bk~ij}^{(V)} &=& \frac{k^{-1}}{2} (\nabla_iY_{\bk~j}^{(V)} +
\nabla_jY_{\bk~i}^{(V)})~.
\eea
Note that contrary to the vector-type vector field $Y_i^{(V)}$, the
vector field $Y_{i}^{(S)}$ is not divergence free. The same is true for
the tensor fields $Y_{ij}^{(S)}$ and  $Y_{ij}^{(V)}$.

Let $T^\mu_\nu=\overline{T}^\mu_\nu+\theta^\mu_\nu$ be the full
energy momentum tensor. We define its energy density $\rho$
and its energy flux 4-vector $u$ as the time-like eigenvalue
and eigenvector of $T^\mu_\nu$:
\be
T^\mu_\nu u^\nu = -\rho u^\mu , \quad u^2=-1.
\ee

We then parameterize their perturbations by
\be
\rho=\bar{\rho}\left(1+\de\right), \quad u=u^0 \dd_t + u^i \dd_i .
\ee
$u^0$ is fixed by the normalization condition,
\be
u^0 = \frac{1}{a} (1-\Psi) .
\ee
We further set
\be
u^i = \frac{1}{a}v^i =\frac{1}{a}\left(VY^{(S)i} + V^{(V)}Y^{(V)i}\right)~.
\ee
Here $\de$ is called the density contrast and $(v^i)$ is the peculiar
velocity. 

We define $P^\mu_\nu\equiv u^\mu u_\nu+\de^\mu_\nu$, the
projection tensor onto the part of tangent space normal to
$u$ and the stress tensor
\be
\tau^{\mu\nu}=P^\mu_\al P^\nu_\beta T^{\al\beta} .
\ee

In the unperturbed case we have
$\tau^0_0=0, \tau^i_j=\bar{p} \de^i_j$. Including
 perturbations, to first order we still obtain
\be
\tau^0_0=\tau^0_i=\tau^i_0=0.
\ee

But $\tau^i_j$ contains in general perturbations. We set
\be
\tau^i_j = \bar{p}\left[\left(1+\pi_L\right)\de^i_j+\Pi^i_j\right]
    , \quad \mbox{with} \quad \Pi^i_i=0.
\ee
We decompose $\Pi^i_j$ into scalar- vector- and tensor-type contributions,
\be
\Pi^i_j =  \Pi^{(S)}Y^{(S)\,i}_j
    + \Pi^{(V)}Y^{(V)\,i}_j
    + \Pi^{(T)} Y^{(T)\,i}_j.
\ee

Another important  variable is
\be
\Ga=\pi_L-\frac{c_s^2}{w}\de
\ee
where $c_s^2\equiv\dot{p}/\dot{\rho}$ is the adiabatic sound speed
and $w\equiv p/\rho$ is the enthalpy. One can show that $\Ga$ is
proportional to the divergence of the entropy flux of the
perturbations. Adiabatic perturbations are characterized
by $\Ga=0$.

We shall use also other  perturbation variables describing the density
contrast and peculiar velocity,
which actually correspond to these perturbations in different
coordinate systems (gauges). One can show that on sub-horizon scales, $k\gg
\HH$, on which perturbations are actually measurable, they all coincide.
\bea
 D &\equiv & \de+3(1+w)
\left(\frac{\da}{a}\right)\frac{V}{k}~,
\label{Ddef}\\
D_g &\equiv& \de - 3(1+w)\Phi ~,  \label{Dgdef}\\[2mm]
\Om &\equiv& V^{(V)} - B^{(V)}~, \\
 \Om - V^{(V)} &=&- B^{(V)} \equiv \si^{(V)} .
\eea
Here we use the customary name, $\si^{(V)}= -B^{(V)}$, for the
vector-type metric perturbation. These variables
 can be  interpreted nicely in terms of gradients
of the energy density and the shear and vorticity of the velocity
field~\cite{Ellis}.

\subsection{Einstein's equations}
\index{Einstein equations}
We do not derive the first order perturbations of Einstein's equations. This can be
done by different methods, for example with Mathematica. We just
write down the results.

\subsubsection{Constraint equations}

\bea \left. \begin{array}{rclc}
4\pi G a^2 \rho D & = & -(k^2 -3K)\Phi & (00)\\
4\pi G a^2 (\rho+p) V & = & k\left(\HH\Psi+ \dot{\Phi}\right) & (0i)
\end{array} \right\} && \mr{(scalar)}\label{scalcntr} \\
8\pi G a^2 (\rho+p)\Om ~~=~~  \frac{1}{2} \left(2K-k^2\right)\si^{(V)}
\quad (0i) \quad && \mr{(vector)} \label{veccntr}
\eea

\subsubsection{Dynamical equations}
\bea
k^2 \left(\Phi-\Psi\right) &=& 8\pi G a^2 p\Pi^{(S)}
\qquad (i\neq j) \quad\quad \mr{(scalar)} \label{scaldyn}\\
\HH\left[ \dot\Psi + \left(\frac{\HH^2- \dot\HH}{\HH^2}\Phi +
   \HH^{-1}\dot\Phi\right)^{\bullet}\right] + && \\   \left(\HH^2
+2\dot\HH\right)\left[\Psi +\frac{\HH^2- \dot\HH}{\HH^2}\Phi +
   \HH^{-1}\dot\Phi\right] &=& 4\pi Ga^2\left(c_s^2 D_g + w\Ga
-\frac{2}{3}w\Pi\right)
\qquad (i~i) \quad \mr{(scalar)} \label{scaldyn2}\\
k\left(\dot{\si}^{(V)}+2\left(\frac{\da}{a}\right)\si^{(V)}\right) &=&
8\pi G a^2 p \Pi^{(V)}
 \qquad (i j) \qquad\qquad \qquad\qquad ~ \mr{(vector)} \label{vecdyn} \\
\ddot{H}^{(T)}
    +2\left(\frac{\da}{a}\right)\dot{H}^{(T)}
    +\left(2K+k^2\right)H^{(T)}  &=&
8\pi G a^2 p \Pi^{(T)}
\qquad (ij) \qquad\qquad\qquad\qquad ~ \mr{(tensor)} \label{tensdyn}
\eea
 For perfect fluids, where $\Pi^i_j\equiv 0$, we have
$\Phi=\Psi$, $\si^{(V)}\propto 1/a^2$, and $H^{(T)}$ obeys a damped
wave equation. The damping term can be neglected on small scales (over
short time periods) when
$\eta^{-2}\lsim 2K+k^2$, so that $H^{(T)}$ represents a propagating
gravitational wave. For vanishing curvature, the scales $k\eta\gg 1$
are simply the sub-horizon scales. For $K<0$, waves
oscillate with a somewhat smaller frequency, $\om=\sqrt{2K+k^2}$,
 while for $K>0$ the frequency is somewhat higher than $k$.

\subsubsection{Energy momentum conservation}
The conservation equations, $\nabla_\nu T^{\mu\nu} \equiv
T^{\mu\nu}_{;\nu}=0$ lead to the following perturbation  equations.
\bea \hspace{-1cm}\left. \begin{array}{c}
\dot{D}_g+3\left(c_s^2-w\right)\left(\frac{\da}{a}\right)D_g
+(1+w)kV+3w\left(\frac{\da}{a}\right)\Ga=0 \\
\dot{V}+\left(\frac{\da}{a}\right)\left(1-3c_s^2\right)V = k\left(\Psi
+ 3c_s^2\Phi\right) + \frac{c_s^2 k}{1+w}D_g \\
 \hspace{2cm} +\frac{wk}{1+w}\left[\Ga-\frac{2}{3}\left(1 -
 \frac{3K}{k^2}\right)\Pi\right]
\end{array} \right\} && \mr{(scalar),} \label{scalcons}\\
\hspace{-1cm}\dot{\Om}+\left(1-3c_s^2\right)\left(\frac{\da}{a}\right)\Om
=\frac{p}{2(\rho+p)}\left(k-\frac{2K}{k}\right)\Pi^{(V)} \qquad &&
\mr{(vector).} \label{veccons}
\eea
These can of course also be obtained from the Einstein equations since they are
equivalent to the contracted Bianchi identities. 

For scalar perturbations we have 4 independent equations and 6
variables. For vector perturbations we have 2 equations and 3
variables, while for tensor perturbations we have 1 equation and 2
variables. To close the system we must add matter equations. The
simplest prescription is to  set $\Ga=\Pi_{ij}=0$. These matter
equations, which describe adiabatic perturbations of a perfect fluid give us
exactly two additional equations for scalar perturbations and one each
for vector and tensor perturbations.

Another example is a universe with matter content given by a
scalar field. We shall discuss this case in the next section. More
complicated examples are those of several interacting particle species of
which some have to be described by a Boltzmann equation. This is the
actual universe at late times, say $z\lsim 10^{10}$.

\subsubsection{A special case}
Here we rewrite the scalar perturbation equations for a simple but
important special case.  We consider adiabatic perturbations of
a perfect fluid. In this case $\Pi=0$
 and $\Ga=0$. Eq.~(\ref{scaldyn}) implies $\Phi=\Psi$. Using
the first equation of (\ref{scalcntr}) and
Eqs. (\ref{Dgdef},\ref{Ddef}) to replace $D_g$ in the second of
Eqs.~(\ref{scalcons}) by $\Psi$ and $V$, finally
replacing $V$  by (\ref{scalcntr}) one can derive a second order
equation for $\Psi$, which is, the only dynamical degree
of freedom
\be \label{Psifluid}
\ddot\Psi + 3\HH(1+c_s^2)\dot\Psi +[(1+3c_s^2)(\HH^2-K) -(1+3w)(\HH^2+K)
  +c_s^2k^2]\Psi = 0~.
\ee

Another interesting example (especially when discussing inflation) is the
scalar field case. There, as we shall see in Section~\ref{scalarfield},
$\Pi=0$, but in general $\Ga\neq 0$ since $\de p/\de\rho\neq \dot
p/\dot\rho$. Nevertheless, since this case again has only one
dynamical degree of freedom, we can express the perturbation equations
in terms of one single second order equation for $\Psi$. In
Section~\ref{scalarfield}
we shall find the following equation for a perturbed scalar field cosmology
\be \label{Psifield}
\ddot\Psi + 3\HH(1+c_s^2)\dot\Psi +[(1+3c_s^2)(\HH^2-K) -(1+3w)(\HH^2+K)
  +k^2]\Psi = 0~.
\ee

The only difference between the perfect fluid and scalar field
perturbation equation is that the latter is missing the factor $c_s^2$
in front of the oscillatory $k^2$ term. Note also that for $K=0$ and
$w=c_s^2=$ constant, the time dependent mass term
$m^2(\eta)=-(1+3c_s^2)(\HH^2-K) + (1+3w)(\HH^2+K)$ vanishes.

It is useful to define the variable~\cite{Mukhanov:1992tc}
\be\label{defu}
 u = a\left[4\pi G(\HH^2-\dot\HH+K)\right]^{-1/2}\Psi,  \ee
which satisfies the equation
\be \label{ueq}
\ddot u +(\Upsilon k^2 -\ddot\theta/\theta)u = 0,
\ee
where $\Up = c_s^2$ or $\Up = 1$ for a perfect fluid or a scalar field
background respectively, and
\be
 \theta = \frac{3\HH}{2a\sqrt{\HH^2-\dot\HH+K}}~.
\ee

Another interesting variable is
\be\label{curva}
\ze \equiv \frac{2(\HH^{-1}\dot\Psi +\Psi)}{ 3(1+w)} +\Psi ~.
\ee
For the rest of this section we set $K=0$ for simplicity. Using
Eqs.~(\ref{Psifluid}) and (\ref{Psifield}) respectively one then finds
\be \label{cons}
\dot\ze = -k^2\frac{\Up\HH}{ \HH^2-\dot\HH}\Psi ~.
\ee
On super-horizon scales, $k/\HH\ll 1$, this time derivative is
suppressed by a factor $\sim (k/\HH)^2 \simeq (k\eta)^2$ and this variable is
(nearly) conserved on large scales.

The evolution of $\ze$ is closely related to the canonical variable
$v$ defined by
\be \label{defv}
v = -\frac{a\sqrt{\HH^2-\dot\HH}}{\sqrt{4\pi G}\Up\HH}\ze~,
\ee
which satisfies the equation
\be
 \ddot v +(\Up k^2- \ddot z/z)v =0~,  
\quad \mbox{ for }    \label{z}
z = \frac{a\sqrt{\HH^2-\dot \HH +\ka}}{ \Up \HH}~.
\ee

\subsection{Dust and radiation}
Next we discuss two simple applications which
are important to understand the  anisotropies in the cosmic microwave
background (CMB).

\subsubsection{The pure dust fluid for $K=0, \La=0$}
'Dust' is the cosmological term for non-relativistic particles for
which we can neglect the pressure so that $w=c_s^2=p=0$ and
$\Pi=\Ga=0$. The Friedmann equation implies for dust $a\propto \eta^2$
so tha $\HH =2/\eta$.
Equation (\ref{Psifluid}) then reduces to
\be \label{Psidust}
\ddot\Psi + \frac{6}{\eta}\dot\Psi  = 0~,
\ee
with the general solution
\be
\label{Psidust2}
\Psi = \Psi_0 +\Psi_1\frac{1}{ \eta^5}
\ee
with arbitrary constants $\Psi_0$ and $\Psi_1$.
Since the perturbations are supposed to be small initially,
they cannot diverge for $\eta \ra 0$, and we have therefore
to keep only the 'growing' mode, $\Psi_1=0$.
 But also the  $\Psi_0$ mode is only constant. This fact led
Lifshitz who was the first to analyze cosmological perturbations to
the conclusions that linear perturbations do not grow in a Friedman
universe and cosmic structure cannot have evolved by gravitational
instability~\cite{Lif46}. However, the important point to note here
is that, even if the gravitational potential remains constant, matter density
fluctuations do grow on sub-horizon scales, scales where $k\eta\gg 1$ and
hence structure can evolve on scales which are smaller
than the Hubble scale.

Defining $x=k\eta$,  we obtain for the velocity potential and the
density contrast
\bea
V &=& \Psi_0 \frac{x}{3}  \label{eq48}\\
D_g &=& -5 \Psi_0 -\frac{1}{6}\Psi_0 x^2~, \qquad
 D =D_g + 3\Psi +\frac{6}{x}V = -\frac{1}{6}\Psi_0 x^2 ~.
\eea
In the variable $D$ the constant term has disappeared and we have $D\ll \Psi$
on super-horizon scales, $x\ll 1$.

On sub-horizon scales, the density fluctuations grow like the scale
factor $a \propto x^2$. Nevertheless, Lifshitz'
conclusion~\cite{Lif46} that pure gravitational
instability cannot be the cause for structure formation has some truth:
if we start from tiny thermal fluctuations of the order
of $10^{-35}$, they can only grow to about
$10^{-30}$ due to this mild, power law instability during the matter dominated
regime. Or, to put it differently,
if we want to form structure by gravitational instability,
we need initial fluctuations of the order of at least $10^{-5}$, much
larger than thermal fluctuations. According to what we have said
here, we need these fluctuations at the beginning of the matter
dominated phase, but as we shall see below, perturbations do not grow at all
during the radiation dominated era, so that really \textit{initial
  fluctuations} with amplitudes $\simeq 10^{-5}$ are needed.
One possibility to create such fluctuations is quantum particle
production in the classical gravitational field during inflation.
 The rapid expansion of the universe during inflation quickly expands
microscopic scales, at which quantum fluctuations are important, to
cosmological scales where these fluctuations are then ``frozen in'' as
classical perturbations in the energy density and the geometry.
We will discuss the induced spectrum on fluctuations in Section~\ref{sec:inf}.

\subsubsection{The pure radiation fluid, $K=0, \La=0$}
\label{radiation}

In this limit we set $w=c_s^2=\nicefrac1/3$, and $\Pi=\Ga=0$ so that
$\Phi=-\Psi$. We conclude from $\rho\propto a^{-4}$ and the Friedmann
equation that $a\propto\eta$.
For radiation, the $u$--equation (\ref{ueq}) becomes
\be \label{ueq2}
\ddot u +(\frac{1}{3} k^2 - \frac{ 2}{ \eta^2})u = 0,
\ee
with general solution
\be \label{urad}
 u(x) = A\left(\frac{\sin(x)}{ x} -\cos(x)\right) + B\left(\frac{\cos(x)}{ x}
 -\sin(x)\right) ~,
\ee
where we have set $x=k\eta/\sqrt{3}=c_sk\eta$. For the Bardeen
potential we obtain with (\ref{defu}), up to constant factors,
\be \label{Psirad}
 \Psi(x) =\frac{u(x)}{ x^2} ~.
\ee
We must set $B=0$ for perturbations to remain regular at early times.
On super-horizon scales, $x\ll 1$, we then have
\be
 \Psi(x) \simeq \frac{A}{ 3}~.
\ee
For the density and velocity perturbations one finds
\be
D_g = 2A \left[\cos(x)   -\frac{2}{ x}\sin(x)\right]  ~, \qquad
V=-\frac{\sqrt{3}}{4}D'_g  ~.
\ee
In the  {\bf super-horizon regime}, $x\ll1$,  this yields
\be
\Psi=\frac{A}{ 3}, \quad D_g=-2A - \frac{A}{ 3\sqrt{3}} x^2, \quad
V=\frac{A}{ 2\sqrt{3}} x~.
\ee
On {\bf sub-horizon scales}, $x\gg1$,  we obtain oscillating
 solutions with constant amplitude and
with frequency  $k/\sqrt{3}$:
\be
V=    \frac{\sqrt{3}A}{ 2} \sin(x)~, \quad
D_g=  2A \cos(x)~,~~~ \Psi=-A\cos(x)/x^{2}~.
\ee
Note that also radiation perturbations outside the Hubble
horizon are frozen to first order. Once they enter the horizon
they start to collapse, but pressure resists the gravitational
force and the radiation fluid fluctuations oscillate at constant
amplitude. The perturbations
of the gravitational potential oscillate and decay like $1/a^2$
inside the horizon.

\subsubsection{ Adiabatic initial conditions}

Adiabaticity requires that the perturbations of all contributions to
the energy density are initially in thermal equilibrium. This fixes
the ratio of the density perturbations of different components. There is no
entropy flux and thus $\Ga=0$. Here we consider a mixture of
non relativistic matter and radiation.
Since the matter and radiation perturbations behave in the
same way on super-horizon scales,
\be
D_g^{(r)}=A+Bx^2, \quad
D_g^{(m)}=A'+B'x^2, \quad
V^{(r)} \propto V^{(m)} \propto x,
\ee
we may require a constant ratio between matter and radiation
perturbations. As we have seen in the previous section,
inside the horizon ($x>1$) radiation
perturbations start to oscillate while matter perturbations keep
following a power law. On sub-horizon scales a constant ratio
can thus no longer be maintained. There are two interesting
possibilities: adiabatic and isocurvature perturbations. Here we
concentrate on adiabatic perturbations which seem to dominate
the observed CMB anisotropies.

From $\Ga=0$ one easily
derives that two components with $p_i/\rho_i =w_i =$constant,
$i=1,2$, are adiabatically coupled if $(1+w_1)D_g^{(2)}=(1+w_2)D_g^{(1)}$.
Energy conservation then  implies that their
velocity fields agree, $V^{(1)}=V^{(2)}$. This result is also a
consequence of the Boltzmann equation in the strong coupling regime.
 We therefore require
\be
V^{(r)}=V^{(m)} ,
\ee
so that the energy flux in the two fluids is coupled initially.

We restrict ourselves to a matter dominated background, the situation
relevant in the observed universe after equality. We first have to
determine the radiation perturbations during a {\em matter dominated
  era}. Since $\Psi$ is dominated by the matter contribution (it is
proportional to the background density of a given component), we
have  $\Psi\simeq \mr{const}=\Psi_0$. We neglect the
contribution from the sub-dominant radiation
to $\Psi$. Energy momentum conservation for radiation then
gives, with $x=k\eta$, and $d/dx =  \prime$
\bea
D_g^{(r)\prime} &=& -\frac{4}{3} V^{(r)}\\
V^{(r)\prime}    &=& 2\Psi + \frac{1}{4} D_g^{(r)} .
\eea

Here $\Psi$ is just a constant given by the matter
perturbations, and it acts like a constant source term.
The general solution of this system is then
\bea
D_g^{(r)} &=& A \cos(c_sx)
    - \frac{4}{\sqrt{3}}B \sin(c_sx)
    + 8 \Psi \left[\cos(c_sx)-1\right]\\
V^{(r)} &=& B \cos(c_sx)
    + \frac{\sqrt{3}}{4}A \sin(c_sx)
    + 2\sqrt{3} \Psi \sin(c_sx) ,
\eea
where $c_s=1/\sqrt{3}$ is the sound speed of radiation.
Our adiabatic initial conditions require
\be
\lim_{x\rightarrow 0} \frac{V^{(r)}}{x}=V_0
=\lim_{x\rightarrow 0} \frac{V^{(m)}}{x} < \infty .
\ee
Therefore $B=0$ and $V=V_0x$ with $V_0=A/4-2\Psi$ on super horizon
scales, $x\ll 1$. Using in addition $\Psi=3V_0$
(see (\ref{eq48})) we obtain
\bea
D_g^{(r)} &=& \frac{4}{3} \Psi \cos\left(\frac{x}{\sqrt{3}}\right)
    - 8 \Psi \label{Dad}\\
V^{(r)} &=& \frac{1}{ \sqrt{3}} \Psi \sin\left(\frac{x}{\sqrt{3}}\right)\\
D_g^{(m)} &=& -\Psi( 5 + \frac{1}{6} x^2) \\
V^{(m)} &=& \frac{1}{ 3} \Psi x .
\eea
On super-horizon scales, $x\ll 1$ we have
\be
    D_g^{(r)}  \simeq -\frac{20}{3} \Psi ~\mbox{ and }~~~ V^{(r)}
    \simeq \frac{1}{3}x \Psi~, \label{Dads}
\ee
note that $D_g^{(r)} = (4/3)D^{(m)}_g$ and $V^{(r)}=V^{(m)}$ as it is
required for adiabatic initial conditions.

\subsection{Scalar field cosmology}
\label{scalarfield}
\index{scalar field}
We now consider the special case of a Friedmann universe filled with
self interacting scalar field matter. We keep spatial curvature $K=0$
in this section. The action is given by
\be
 S = \frac{1}{ 16\pi G}\int d^4x\sqrt{|g|}R  +
  \int d^4x\sqrt{|g|}\left(\frac{1}{ 2}\dd_\mu\vph\dd^\mu\vph - W(\vph)\right)
 = S_g + S_m
\ee
where $\vph$ denotes the scalar field and $W$ is the potential. The
energy momentum tensor is obtained by varying the matter part of the
action, $S_m$ wrt the metric $g^{\mu\nu}$,
 \be
   T_{\mu\nu} = \dd_\mu\vph \dd_\nu\vph -
   \left[\frac{1}{ 2}\dd_\la\vph\dd^\la\vph + W \right]g_{\mu\nu}
\ee
The energy density $\rho$ and the energy flux $u$ are defined by
\be
 T^\mu_\nu u^\nu = -\rho u^\mu ~.
\ee
For a homogeneous and isotropic universe, $\vph=\vph(t)$ and
$g_{\mu\nu} = a^2\eta_{\mu\nu}$ we obtain
\be \rho = \frac{1}{ 2a^2}\dot\vph^2 + W  ~~~~~~~~ (u^\mu) = \frac{1}{
a}( 1, {\bf 0}) ~. \ee
The pressure is given by
\be T^i_j = p\de^i_j  \qquad  p = \frac{1}{ 2a^2}\dot\vph^2 - W  ~. \ee

We now consider scalar field perturbations,
\be  \label{defphi}
\vph = \bar{\vph} +\de\vph ~.  \ee
Clearly, the scalar field only generates scalar-type perturbations (to
first order).
The perturbed metric is therefore given by $ds^2=-a^2(1+2\Psi)d\eta^2
+ a^2(1-2\Phi)\de_{ij}dx^idx^j$.
Inserting Eq.~(\ref{defphi}) in the definition of the energy velocity
perturbation $V$,
\be (u^\mu) = \frac{1}{ a}( 1-\Psi, -V,_i) \ee
and the energy density perturbation $\de\rho$,
\be \rho = \bar{\rho} +\de\rho ~, \ee
we obtain
\be \de\rho  =\frac{1}{ a^2}\dot{\bar{\vph}}\de\dot{\vph} - \frac{1}{
a^2}\dot{\bar{\vph}}^2\Psi + W,_\vph\de\vph  \ee
and
\be V = \frac{k}{ \dot{\bar{\vph}}}\de\vph  ~. \ee
From the stress tensor, $T_{ij} = \vph,_i  \vph,_j -
   \left[\frac{1}{ 2}\dd_\la\vph\dd^\la\vph + W \right]g_{ij}$ we find
\be p\pi_L =  \frac{1}{ a^2}\dot{\bar{\vph}}\de\dot{\vph} - \frac{1}{
a^2}\dot{\bar{\vph}}^2\Psi - W,_\vph\de\vph \qquad \mbox{and }\quad \Pi = 0 ~.\ee
 Short calculations give
\begin{eqnarray}
D_g &=& -(1+w)\left[4\Psi + 2\frac{\dot a}{ a}k^{-1}V- k^{-1}\dot V
  \right] ~, \\
D_s &=& D_g+3(1+w)\Psi ~,\\
\Ga &=& \frac{2W,_\vph}{ p\dot\rho}\left[\dot{\bar{\vph}}\rho D_s -
  \dot\rho \de\vph\right] ~, \\  \Pi &=& 0~.
\end{eqnarray}
The Einstein equations then lead to the following second order
equation for the Bardeen potential which we have discussed above:
\be  \ddot\Psi + 2({\HH}-\ddot\vph/\dot\vph)\dot\Psi
+(2\dot{\HH}-2{\HH}\ddot\vph/\dot\vph +k^2)\Psi =0
 \ee
or, using the definition $c_s^2= \dot p/\dot\rho$,
\be  \ddot\Psi + 3{\HH}(1+c_s^2)\dot\Psi +(2\dot{\HH}
+(1+3c_s^2){\HH}^2 +k^2)\Psi =0~.
 \ee
 As already mentioned above, this equation differs from the $\Psi$
 equation for a perfect fluid only in the last term proportional to
 $k^2$. This comes from the fact that the scalar field is not in a
 thermal state with fixed entropy, but it is in a fully coherent state
 ($\Ga\neq 0$) and field fluctuations propagate with the speed of
 light. On large scales, $k\eta\ll 1$, this difference is not relevant,
 but on sub--horizon scales it does play a certain role.

\subsection{Generation of perturbations during inflation}
\index{inflation}
\label{sec:inf}
 So far we have simply assumed some initial fluctuation amplitude $A$,
without investigating where it came from or what the $k$--dependence of $A$
might be. In this section we discuss the most common idea about
the generation of cosmological perturbations, namely their production
from quantum vacuum fluctuations during an inflationary phase. The
treatment here is focused  mainly on getting the correct result with
as little effort as possible; we ignore several subtleties related,
\eg to the transition from quantum fluctuations of the field to
classical fluctuations in the energy momentum tensor. The idea is of
course that the source for the  metric fluctuations are the {\em expectation
values} of the energy momentum tensor operator of the scalar field.

The basic idea is simple: A time dependent gravitational field
very generically leads to particle production, analogously to the
electron positron production in a classical, time dependent,
strong electromagnetic field.

Let us first fix our notation. Inflation is an era during which the
expansion of the scale factor is accelerated, $\frac{d^2
  a}{dt^2}>0$. In terms of conformal time, $\frac{d}{d\eta} = \dot{~}$~,
this becomes
$$
\frac{d^2 a}{dt^2} = \frac{1}{a}\dot\HH >0~.
$$
We shall only consider simple power law inflation, where $a=(c\eta)^q$
for some constants $c$ and $q$. For the scale factor to be positive
and real we require $c\eta>0$. Expansion then happens when $cq>0$ and
accelerated expansion when in addition $q<0$. Hence for power law
inflation, the scale factor behaves like
$$  a \propto |\eta|^q $$
and $\eta<0$ as well as $q<0$. It is easy to see that de Sitter
inflation, $a\propto \exp(Ht)$, corresponds to $q=-1$. In general, for
a fluid with $p=w\rho$
$$ q= \frac{2}{1-3w} ~.$$
Inflation therefore requires $w<-1/3$.
During scalar field inflation, the energy density must therefore be
dominated by the potential, $W>a^{-2}\dot\vph^2$. We suppose that the
field is 'slowly 
rolling' down the potential until at some later moment the condition
 $w<-1/3$ breaks down and inflation stops. How far away a given moment
is from this end of inflation can be cast in terms of the slow roll parameters
$\ep_1$ and $\ep_2$ defined by
\bea
\ep_1 &=& -\frac{\dot H}{aH^2} ~, \quad H =\frac{\HH}{a} \quad \mbox{
  is the Hubble parameter}\\
\ep_2 &=& -\frac{\frac{a^2d^2\vph}{dt^2}}{\HH\dot\vph}
  = \left[1 -\frac{\ddot{\vph}}{\HH\dot\vph}\right] =
\left[1 +\frac{a^2W'}{\HH\dot\vph}\right] ~ .
\eea

\subsubsection{Scalar perturbations}
The main result of this subsection is the following: During inflation,
the produced particles induce a  gravitational field with a
(nearly) scale invariant spectrum,
\be  \label{specin}
k^3|\Psi(k,\eta)|^2 = k^{n-1}\times\mr{const.} \quad \mbox{ with }
 \quad n\simeq 1~.
\ee

The quantity $k^3|\Psi(k,\eta)|^2$ is the squared amplitude of the
metric perturbation at comoving scale $\la=\pi/k$. To ensure
that this quantity is small over a broad range of scales, so that
neither black holes form on small scales nor large
deviation from homogeneity and isotropy on large scales appear, we must
require $n\simeq 1$. These arguments have been put forward
 by Harrison and Zel'dovich~\cite{HZ} (ten years before the advent of
inflation), leading to the name
'Harrison-Zel'dovich spectrum' for a scale invariant perturbation
spectrum.

To derive the above result we consider a scalar field background dominated
by a potential, hence $a\propto |\eta|^{q}$ with $q\sim -1$.
Developing  the action of this system,
\[ S = \int dx^4\sqrt{|g|}\left(
   \frac{R}{ 16\pi G} +\frac{1}{ 2}(\nabla\vph)^2 - W \right)~,
\]
to second order in the perturbations (see~\cite{Mukhanov:1992tc})
around the Friedmann solution one obtains
\be \label{action}
 \de S =  \int dx^4\sqrt{|\overline g|}\frac{1}{ 2}(\dd_\mu v)^2
\ee
up to a total differential. Here $v$ is the perturbation variable
\be v = -\frac{a\sqrt{\HH^2-\dot\HH}}{ \sqrt{4\pi G}\HH}\ze \ee
introduced in Eq.~(\ref{defv}). Via the Einstein equations, this
variable can also be interpreted as representing the fluctuations in
the scalar field. Therefore, we quantize $v$ and assume that
initially, on small scales, $k|\eta|\ll 1$, $v$ is in the (Minkowski) quantum
vacuum state of a massless scalar field with mode function
\be \label{vin}
  v_\mr{in} = \frac{v_0}{ \sqrt{k}}\exp(ik\eta) ~.
\ee
The pre-factor $v_0$ is a $k$-independent constant which depends on
convention, but is of order unity.
From (\ref{cons}) we can derive
\[ (v/z)^{\textstyle\cdot} = \frac{k^2 u}{ z} ~,\]
where $z\propto a$ is defined in Eq.~(\ref{z}) and $u \propto
a\eta\Psi$ is given in Eq.~(\ref{defu}). On small scales, $k|\eta|\ll 1$,
this results in the initial condition for $u$
\be\label{uin} u_\mr{in} = \frac{-iv_0}{ k^{3/2}}\exp(ik\eta) ~.\ee

In the case of power law expansion, $a\propto |\eta|^q$, the evolution
equation for $u$, Eq.~(\ref{ueq}), reduces  to
\be
\ddot u +( k^2 -\frac{q(q+1)}{ \eta^2})u = 0.
\ee
The solutions to this equation are of the form
$(k|\eta|)^{1/2}H^{(i)}_\mu(k\eta)$, where $\mu=q+1/2$ and
$H^{(i)}_\mu$ is the Hankel function of the $i$th kind ($i=1$ or $2$) of order
$\mu$. The initial condition (\ref{uin}) requires that only $H^{(2)}_\mu$
appears, so that we obtain
\[
 u = \frac{\al}{ k^{3/2}}(k|\eta|)^{1/2}H^{(2)}_\mu(k\eta) ~,
\]
where again $\al$ is a constant of order unity. We define the value of
the Hubble parameter during inflation, which is nearly constant by
$H_i$. With $H=\HH/a \simeq 1/(|\eta| a)$, we then obtain  $a\sim
1/(H_i|\eta|)$. With the
Planck mass defined by  $4\pi G=M^{-2}_4$,  Eq.~(\ref{defu}) then gives
\be
\Psi = \frac{H_i}{ 2M_{4}}u \simeq
  \frac{H_i}{ M_{4}} k^{-3/2}(k|\eta|)^{1/2}H^{(2)}_\mu(k\eta) ~.
\ee
 On small scales this is a simple oscillating function
while on large scales $k|\eta|\ll1$ it can be approximated by a
power law,
\be \Psi  \simeq  \frac{H_i}{ M_{4}}
k^{-3/2}(k|\eta|)^{1+q}
   ~,~\mbox{ for }~ k|\eta|\ll 1~~~~.
\ee
Here we have used $\mu=1/2+q<0$.  This yields
\be\label{infspec}
k^3|\Psi|^2 \simeq  \left( \frac{H_i}{ M_{4}}(k|\eta|)^{1+q}\right)^2
\propto k^{n-1} ~,
\ee
hence $n\simeq 1$  if $q\sim -1$.
 Detailed studies have shown that  the
amplitude of $\Psi$ can still be somewhat affected by the transition
from inflation to the subsequent radiation era, the spectral index, however,
is very stable. Simple deviations from de Sitter inflation, like \eg
power law inflation, $q>-1$, lead to
slightly blue spectra, $n\gsim 1$.

With a somewhat more careful treatment, one finds that both, the
amplitude and the spectral index depend on scale via the slow roll parameters
$\ep_1$ and $\ep_2$,
\bea  \label{e:srPsi}
   k^3|\Psi|^2 &=& \frac{2H_i^2}{M_4^2\ep_1}(k\eta_f)^{n-1} ~,\\
\left. n\right|_{k=a(\eta)H(\eta)} &=& 1-4\ep_1(\eta)- 2\ep_2(\eta) ~.
\eea

Vector perturbations are not generated during standard inflation; and
even if they are generated they only decay during subsequent
evolution and we therefore do not discuss them any further. This may
change drastically in braneworlds (see \cite{Ringeval:2003na})!

\subsubsection{Tensor perturbations}

The situation is different for tensor perturbations. Again we
consider the perfect fluid case, $\Pi_{ij}^{(T)}=0$. Eq.
(\ref{tensdyn}) implies
\be
\ddot H_{ij}+\frac{2\dot a}{a}\dot H_{ij}+k^2H_{ij}=0~. \label{tens}
\ee
 If the background has a power law evolution, $a\propto \eta^q$ this
 equation can be solved in terms of Bessel or Hankel functions. The
 less decaying  mode solution to Eq.~(\ref{tens}) is
$H_{ij} =e_{ij}x^{1/2-\beta}J_{1/2-q}(x)$, where $J_\nu$ denotes the
Bessel function of order $\nu$, $x=k\eta$  and $e_{ij}$ is a transverse traceless
polarization tensor. This leads to
\bea
H_{ij}&=&\mr{const}\quad\mr{for}\quad x\ll1\\
H_{ij}&=&\frac{1}{a}\quad\mr{for}\quad x\gsim 1 ~.
\eea
One may also quantize the tensor fluctuations which represent
gravitons. Doing this, one obtains (up to log corrections) a
scale invariant spectrum of tensor fluctuations from inflation:
for tensor perturbations the canonical variable is simple given by
$h_{ij} = M_PaH_{ij}$.
The evolution equation for $h_{ij} =he_{ij}$ is of the form
\be \ddot h +(k^2+m^2(\eta)) h =0 ~,\label{tencan} \ee
where $m^2(\eta)=-\ddot a/a$. During inflation $m^2=-q(q-1)/\eta^2$ is
negative, leading to particle creation. Like for scalar perturbations,
the vacuum initial conditions are given on scales
which are inside the horizon, $k^2\gg |m^2|$,
\[h_\mr{in} = \frac{1}{\sqrt{k}}\exp(ik\eta) ~~\mbox{ for }~ k|\eta|\gg  1.\]
Solving Eq.~(\ref{tencan}) with this initial condition, gives
\[ h = \frac{1}{ \sqrt{k}}(k|\eta|)^{1/2}H^{(2)}_{q-1/2}(k\eta)~,\]
where $H^{(2)}_\nu$ is the Hankel function of degree $\nu$ of the second kind.
On super-horizon scales where we have $H^{(2)}_{q-1/2}(k\eta)\propto
(k|\eta|)^{q-1/2}$, this  leads to $|h|^2 \simeq
|\eta|(k|\eta|)^{2q-1}$. Using the 
relation between $h_{ij}=he_{ij}$ and $H_{ij}$ one obtains the
spectrum of tensor perturbations generated during inflation. For exponential
inflation, $q\simeq -1$ one finds again a scale invariant spectrum for
$H_{ij}$ on super-horizon scales,
\be
 k^3|H_{ij}H^{ij}| \simeq (2H_\mr{in}/M_4)^2(k\eta_f)^{n_T} \quad
\mbox{ with }\quad  n_T =2(q+1) \simeq 0~.
\ee
Again, a more careful treatment within the slow roll approximation
gives
\be
n_T = -2\ep_1~.
\ee
A more detailed analysis also of the amplitudes of the
scalar and tensor spectra leads to
 the consistency relation $n_T = -2A_T^2/A_S^2$ of slow roll
inflation. Here $A_T$ and $A_S$ are the amplitudes of tensor and scalar
perturbations, respectively.

More details on inflation can be found in many cosmology books, \eg
Refs.~\cite{Dod,LiLy,Linde}. 

\subsection{Power spectra}
\index{power spectra}

The quantities which we have calculated in the previous subsection are
not the precise values of \eg $\Psi(\bk,\eta)$, but only expectation
values $\langle|\Psi(\bk,\eta)|^2\rangle$. In different realizations
of the same inflationary model, the 'phases' $\al(\bk,\eta)$ given by
$\Psi(\bk,\eta) = \exp(i\al(\bk))|\Psi(k)|$ are different. They are
random variables. Since the process which generates the fluctuations
$\Psi$ is stochastically homogeneous and isotropic, these phases are
uncorrelated (for different values of $\bk$). However, the quantity
which we can calculate for a given model and which then has to be
compared with observation is the power spectrum.  Power spectra are
the ``harmonic transforms'' of the two point correlation
functions\footnote{The ``harmonic transform'' in usual flat space is
  simply the Fourier transform. In curved space it is the expansion in
  terms of eigenfunctions of the Laplacian on that space, \eg on the
  sphere it corresponds to the expansion in terms of spherical
  harmonics}. 
If the perturbations 
of the model under consideration are Gaussian, this is a relatively generic
prediction from inflationary models, then the two-point functions and
therefore the power spectra
contain the full statistical information of the model.

Let us first consider the power spectrum of matter,
\be
P_D(k)=\left\lan\left| D_g\left(\bk,\eta_0\right)\right|^2\right\ran ~.
\ee
Here $\lan~\ran$ indicates a statistical average, ensemble average,
 over ``initial
conditions'' in a given model. $P_D(k)$ is usually compared with
the observed power spectrum of the galaxy distribution.

The spectrum we can both, measure and calculate to the best
accuracy is the CMB anisotropy power spectrum. It is defined as follows:
The fluctuations of the radiation temperature as observed in the
sky, $\De T/T$, is a function of our position $\bx_0$, time $\eta_0$ and
the photon direction $\bn$. We develop the $\bn$--dependence in
terms of spherical harmonics. We will suppress the argument $\eta_0$
and often also $\bx_0$ in the
following calculations. All results are for today ($\eta_0$) and here
($\bx_0$). By statistical homogeneity statistical averages over an
ensemble of realizations (expectation values) are supposed
to be independent of position. Furthermore, we assume that the process
generating the initial perturbations is statistically isotropic. Then,
the off-diagonal correlators of the expansion coefficients $a_{\ell m}$
vanish and we have
\be
\frac{\De T}{T}\left(\bx_0,\eta_0,\bn\right)
=\sum_{\ell,m} a_{\ell m}(\bx_0,\eta_0) Y_{\ell m}(\bn), \quad
\left\lan a_{\ell m}\cdot a_{\ell'm'}^*\right\ran
= \de_{\ell\ell'}\de_{mm'}C_\ell ~.
\ee
The $C_\ell$'s are the CMB power spectrum.

The two point correlation function is related to the $C_\ell$'s by
\bea
\left\lan \frac{\De T}{T}(\bn)\frac{\De T}{T}(\bn')\right\ran_{\bn\cdot\bn'=\mu}
=  \sum_{\ell,\ell',m,m'}\left\lan a_{\ell m}\cdot a_{\ell'm'}^*\right\ran
     Y_{\ell m}(\bn) Y_{\ell' m'}^*(\bn') =  && \nonumber \\
\sum_\ell C_\ell \underbrace{\sum_{m=-\ell}^\ell
   Y_{\ell m}(\bn) Y_{\ell m}^*(\bn')}_{\frac{2\ell+1}{4\pi}
   P_\ell(\bn\cdot\bn')}
= \frac{1}{4\pi}\sum_\ell (2\ell+1)C_\ell P_\ell(\mu) ,\qquad && \label{correl}
\eea
where we have used the addition theorem of spherical harmonics
for the last equality;  the $P_\ell$'s are the Legendre polynomials.

For given metric perturbations and perturbations of the energy
momentum tensor of the cosmic fluid, the temperature perturbations can
be determined by following the oscillations in the radiation fluid
before  decoupling (see subsection~\ref{radiation}) and by following 
the propagation of photons along geodesics in the perturbed spacetime
after decoupling. Decoupling of photons and matter happens during
recombination ($T\simeq 3000$K, $z\simeq 1000$), where electrons and
protons recombine to neutral hydrogen. During that process, the number
density of free electrons with which the photons can scatter drops
drastically and finally becomes so low, that the
mean free path of the photons grows larger than the Hubble scale. The
surface of constant temperature, $T_\mr{dec}=T(\eta_\mr{dec})$, at
which this happens is also called 
the 'last scattering surface'. After last scattering,  the photons
effectively cease 
to interact and move freely along geodesics (more details can be found
\eg in~\cite{Dod,Rfund,myCMB}).

Clearly the $a_{lm}$'s from scalar-, vector- and tensor-type perturbations
are uncorrelated,
\be
\left\lan a_{\ell m}^{(S)} a_{\ell'm'}^{(V)} \right\ran
=\left\lan a_{\ell m}^{(S)} a_{\ell'm'}^{(T)} \right\ran
=\left\lan a_{\ell m}^{(V)} a_{\ell'm'}^{(T)} \right\ran
=0 .
\ee

Since vector perturbations decay, their contributions, the $C_\ell^{(V)}$,
are negligible in models where initial perturbations have been
laid down very early, \eg, after an inflationary period. Tensor
perturbations are constant on super-horizon scales and perform damped
oscillations once they enter the horizon.

Let us first discuss in somewhat more detail scalar perturbations.
We restrict ourselves to the case $K=0$ for simplicity.
We suppose the initial perturbations to be given by a spectrum,
\be
\left\lan\left|\Psi\right|^2\right\ran k^3=A^2 k^{n-1}\eta_0^{n-1} . \label{inspec}
\ee
We multiply by the constant $\eta_0^{n-1}$, the actual comoving size of the
horizon, in order to keep $A$
dimensionless for all values of $n$. $A$ then represents the amplitude of
metric perturbations at horizon scale today, $k=1/\eta_0$.

 For {\em adiabatic} perturbations we have obtained on {\em
   super-horizon scales}, 
\be  \label{e:Dgr}
\frac{1}{4}D_g^{(r)} = -\frac{5}{3} \Psi +\OO((k\eta)^2), \quad
V^{(b)}=V^{(r)}=\OO(k\eta)~.
\ee

The dominant contribution to the temperature fluctuations on
super-horizon scales (neglecting the 
integrated Sachs--Wolfe effect $\int \dot{\Phi}-\dot{\Psi}$~)
comes from two terms: the first, $2\Psi$, is the change of photon
energy due to the gravitational potential at the last scattering
surface, $\eta=\eta_\mr{dec}$, and the second, 
$\nicefrac1/4D_g^{(r)}$ represents the
intrinsic temperature fluctuations (for more details
see~\cite{SW,peebles,myCMB}). With Eq.~(\ref{e:Dgr}) this yields the
famous Sachs-Wolfe formula
\be
\frac{\De T}{T}(\bx_0,\bn,\eta_0) = 2\Psi(x_\mr{dec},\eta_\mr{dec}) +
\frac{1}{4}D_g^{(r)}(x_\mr{dec},\eta_\mr{dec}) = 
\frac{1}{3} \Psi(x_\mr{dec},\eta_\mr{dec}).     \label{sw}
\ee

The Fourier transform of (\ref{sw}) gives
\be
\frac{\De T}{T}(\bk,\bn,\eta_0)  = \frac{1}{3} \Psi(k, \eta_\mr{dec}) \cdot
    e^{i\bk\bn\left(\eta_0-\eta_\mr{dec}\right)}~.
\ee

Using the decomposition
\[
e^{i\bk\bn\left(\eta_0-\eta_\mr{dec}\right)} =
   {\sum_{\ell=0}^\infty (2\ell+1)i^\ell
   j_\ell(k(\eta_0-\eta_\mr{dec})) P_\ell( \widehat{\bk}\cd\bn)}~,
\]
where $j_\ell$ are the spherical
Bessel functions, we obtain
\bea
\lefteqn{\left\lan \frac{\De T}{T}(\bx_0,\bn,\eta_0)
 \frac{\De T}{T}(\bx_0,\bn',\eta_0) \right\ran}\\
 &=& \frac{1}{V} \int d^3x_0 \left\lan\frac{\De T}{T}(\bx_0,\bn,\eta_0)
 \frac{\De T}{T}(\bx_0,\bn',\eta_0) \right\ran \nonumber \\
&=&\frac{1}{(2\pi)^3}\int d^3k \left\lan\frac{\De T}{T}(\bk,\bn,\eta_0)
 \left(\frac{\De T}{T}\right)^*(\bk,\bn',\eta_0) \right\ran  \nonumber\\
& =& \frac{1}{(2\pi)^3 9}\int d^3k \left\lan\left|\Psi\right|^2\right\ran
    \sum_{\ell,\ell'=0}^\infty
 (2\ell+1)(2\ell'+1) i^{\ell-\ell'} \nonumber \\ &&  \cdot
 j_\ell(k(\eta_0-\eta_\mr{dec}))
 j_{\ell'}(k(\eta_0-\eta_\mr{dec}))P_\ell(\hat{\bk}\cd\bn) \cdot
P_{\ell'}(\hat{\bk}\cd\bn')~.
\eea
In the second equal sign we have used the unitarity of the Fourier
transformation. Inserting $P_\ell(\hat{\bk}\bn) =
 \frac{4\pi}{2\ell+1} \sum_m Y_{\ell m}^*(\hat{\bk})Y_{\ell m}(\bn)$
and    $P_{\ell'}(\hat{\bk}\bn')=
 \frac{4\pi}{2\ell'+1}\sum_{m'} Y_{\ell' m'}^*(\hat{\bk})Y_{\ell' m'}(\bn')$,
the integration over directions $d\Om_{\hat{k}}$
 gives $\de_{\ell\ell'}\de_{mm'}\sum_m Y_{\ell m}^*(\bn)Y_{\ell m}(\bn')$.\\
Using as well $\sum_m Y_{\ell m}^*(\bn)Y_{\ell m}(\bn')=\frac{2\ell+1}{4\pi}
P_\ell(\mu)$, where  $\mu=\bn\cdot\bn'$, we find
\bea
\lefteqn{\left\lan \frac{\De T}{T}(\bx_0,\bn,\eta_0)
 \frac{\De T}{T}(\bx_0,\bn',\eta_0) \right\ran_{\bn\bn'=\mu}
 =} \nonumber \\ &&
\sum_\ell \frac{2\ell+1}{4\pi} P_\ell(\mu) \frac{2}{\pi}
\int\frac{dk}{k} \left\lan\frac{1}{9}|\Psi|^2\right\ran k^3
j_\ell^2(k(\eta_0-\eta_\mr{dec})) .
\eea

Comparing this equation with~Eq.~(\ref{correl})
we obtain for {\em adiabatic perturbations}
on scales $2\le \ell$ $\ll$
$(\eta_0-\eta_\mr{dec})/\eta_\mr{dec}$ $\sim 100$
\be
C_\ell^{(SW)} \simeq 
 \frac{2}{\pi} \int_0^\infty \frac{dk}{k}
 \left\lan\left|\frac{1}{3}\Psi\right|^2
 \right\ran k^3 j_\ell^2 \left(k\left(\eta_0-\eta_\mr{dec}\right)\right).
\label{clsw}
\ee

If $\Psi$ is a pure power law as in Eq.~(\ref{inspec}) and we set
$k(\eta_0-\eta_\mr{dec})\sim k\eta_0$,  the integral (\ref{clsw}) can
be performed analytically. For the ansatz (\ref{inspec}) one finds
\be
C_\ell^{(SW)} = \frac{A^2}{9} \frac{\Ga(3-n)\Ga(\ell-\frac{1}{2}+\frac{n}{2})}{
2^{3-n}\Ga^2(2-\frac{n}{2})\Ga(\ell+\frac{5}{2}-\frac{n}{2})}
\quad\mbox{ for }  -3<n<3~ . \label{clswsol}
\ee

Of special interest is the {\em scale invariant} or
Harrison--Zel'dovich spectrum, $n=1$ (see Section~\ref{sec:inf}).  It leads to

\be
\ell(\ell+1)C_\ell^{(SW)} = \mr{const.} \simeq
\left\lan\left(\frac{\De T}{T}(\vth_\ell)\right)^2\right\ran~,~~~~
    \vth_\ell\equiv \pi/\ell~.
\ee
This  is precisely (within the accuracy of the experiment) the
behavior  observed by the DMR experiment aboard the satellite
COBE \cite{DMR} and more recently with the WMAP satellite~\cite{WMAP}.

Inflationary models predict very generically a HZ spectrum (up to
small corrections). The DMR discovery has therefore been
regarded as a great success, if not a proof, of inflation.
There are other models like topological defects \cite{report}
or certain string
cosmology models \cite{dgsv} which also predict scale--invariant,
\ie Harrison Zel'dovich spectra of fluctuations. These models do
however not belong to the class investigated here, since in these
models perturbations are induced by seeds which evolve non--linearly
in time.

For gravitational waves (tensor fluctuations), a formula analogous to
(\ref{clswsol}) can be derived,
\be
C_\ell^{(T)}=\frac{2}{\pi}\int dk~ k^2 \left\lan\left|\int_{\eta_\dec}^{\eta_0}
d\eta \dot{H}(\eta,k) \frac{j_\ell(k(\eta_0-\eta))}{(k(\eta_0-\eta))^2}
\right|^2\right\ran\frac{(\ell+2)!}{(\ell-2)!} .
\ee

\begin{figure}[ht]
\centerline{\epsfig{figure=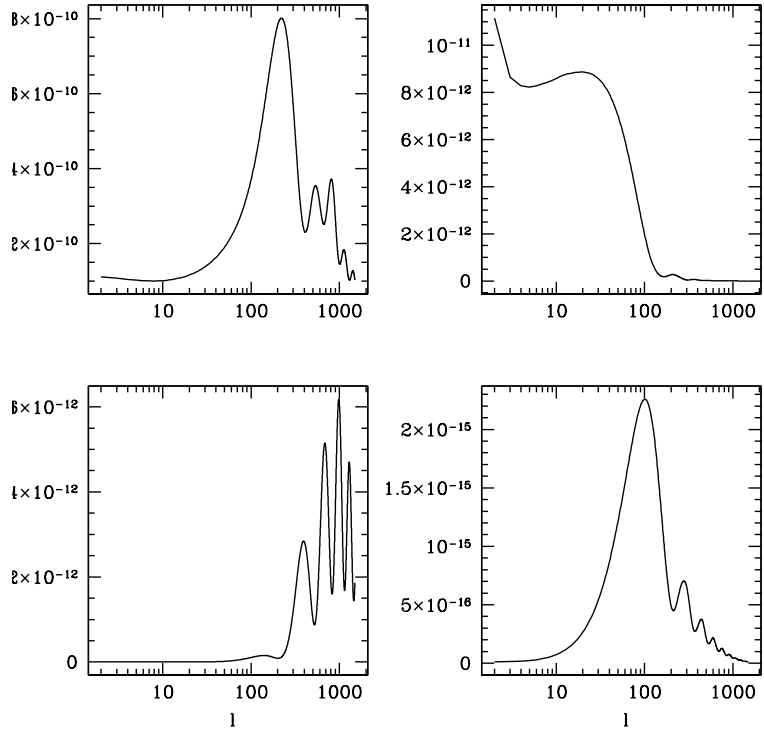, width=10.5cm}}
\caption{\label{STfig}Adiabatic scalar (left) and tensor (right)
 CMB anisotropy spectra are shown. The dimensionless quantity 
$\ell(\ell+1)C_\ell/(2\pi)$ is plotted. The top panels show the
 temerature anisotropies while the
bottom panels show the corresponding polarization spectra (for an
 introduction to polarization see \eg
\cite{Dod}).}
\end{figure}

To a very crude approximation we may assume $\dot{H}=0$ on super-horizon
scales and $\int d\eta \dot{H}j_\ell(k(\eta_0-\eta))
    \sim H(\eta=1/k)j_\ell(k\eta_0)$. For a pure power
law,
\be
k^3\left\lan\left|H(k,\eta=1/k)\right|^2\right\ran \simeq
   A_T^2 k^{n_T}\eta_0^{-n_T} ,
\ee
one obtains
\bea
C_\ell^{(T)} &\simeq & \frac{2}{\pi} \frac{(\ell+2)!}{(\ell-2)!} A_T^2
\int \frac{dx}{x} x^{n_T} \frac{j_\ell^2(x)}{x^4} \nonumber \\
&=& \frac{(\ell+2)!}{(\ell-2)!} A_T^2
    \frac{\Ga(6-n_T)\Ga(\ell-2+\frac{n_T}{2})}{
    2^{6-n_T}\Ga^2(\frac{7}{2}-n_T)\Ga(\ell+4-\frac{n_T}{2})} .
\eea
For a scale invariant spectrum ($n_T=0$) this results in
\be
\ell(\ell+1)C_\ell^{(T)} \simeq \frac{\ell(\ell+1)}{(\ell+3)(\ell-2)}
    A_T^2\frac{8}{ 15\pi}~ .
\ee
The singularity at $\ell=2$ in this crude approximation is not real,
but there is some enhancement of $\ell(\ell+1)C_\ell^{(T)}$
at $\ell\sim2$ see Fig.~\ref{STfig}).
Again, inflationary models (and topological defects) predict a scale
invariant spectrum of tensor fluctuations ($n_T \sim 0$).

On intermediate scales, $100<\ell<1000$, the acoustic oscillations of
radiation density fluctuations before decoupling (see
subsection~\ref{radiation}) lead to a characteristic series of
peaks in the CMB power spectrum which is being measured in great
detail and contains very important information on cosmological
parameters~\cite{JL}.
On small angular scales, $\ell\gsim 800$, fluctuations are damped
by collisional damping (Silk damping). This effect has to be discussed
with the Boltzmann equation for photons, which goes beyond the scope
of this introduction (see~\cite{Dod,JL}).


\section{Braneworld cosmology}\label{sec:cos}
\begin{figure}[ht]
\centerline{\epsfig{figure=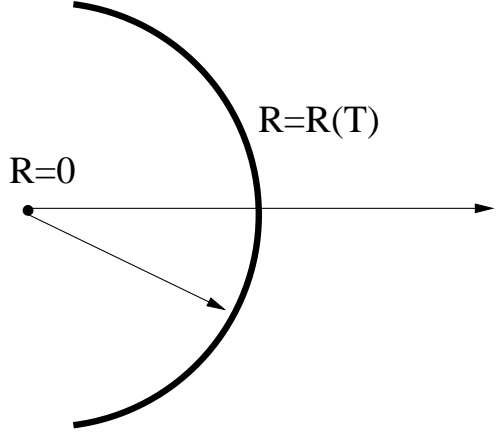, width=6.5cm}}
\caption{A  moving brane in AdS-Schwarzschild. In this coordinates the
  AdS horizon is at $R=\infty$.}
\label{f:branecos}
\end{figure}

We now want to study  cosmology of an expanding
maximally symmetric braneworld. We still require the bulk to be empty
and $Z_2$--symmetric. One can show that the most general empty bulk
allowing for a homogeneous and isotropic brane is Schwarzschild-AdS
(Sch-AdS). In the cosmological setting we allow the brane to move
in the bulk. As we shall see, this can mimic cosmological
expansion. The  situation is as depicted in Fig~\ref{f:branecos}.

The 5d metric of Sch-AdS is of the form
\be
ds^2 = -F(R)dT^2 +\frac{dR^2}{F(R)} + R^2\left(
 \frac{dr^2}{1- Kr^2} + r^2d\Omega^2\right)
\ee
where the function $F$ is determined by the AdS curvature radius $\ell$,
 the 5d mass $C$ and the curvature $K$ of 3d space (on the brane) via
$$   F(R) = K + \frac{R^2}{\ell^2} - \frac{C}{R^2}~.$$

From this we can calculate the 5d Weyl tensor, $C_{ABCD}$, and its
'electric' components defined in (\ref{e:Weylel}) with the result
\bea
E_{\mu\nu} &=& \rho^E u^E_\mu u^E_\nu + \pi^E_{\mu\nu} \qquad
\mbox{with }   u= R^{-1}(1,{\bf 0})~,\\
E_{00} &=& \frac{C}{a^4} = \rho^E \qquad \mbox{and } \qquad
\pi^E_{\mu\nu} = 0~.
\eea
Here $R(T)= a(t)$, where  $t$ is cosmic time on the brane and $u$ is
the unit normal vector in direction of cosmic time on the
brane. $R(T)$ is the brane position at time $t(T)$.

\subsection{The modified Friedmann equations}
From the brane gravity equations,    $$G_{\mu\nu} = -\La_5 g_{\mu\nu} +
\kappa_4 \tau_{\mu\nu} + \kappa_5^2\si_{\mu\nu} - E_{\mu\nu}$$
with $\La_4 = \frac{1}{2}(\La_5 + \kappa_5/6\la^2)$,~~
   $\kappa_4 = \kappa_5^2\la/6$ and
$$
\si_{\mu\nu}  = - \frac{1}{4}\tau_{\mu\alpha}\tau^\alpha_\nu +
\frac{1}{12} \tau\tau_{\mu\nu} -  \frac{1}{8} g_{\mu\nu}
\tau_{\alpha\beta}\tau^{\alpha\beta} - \frac{1}{24}g_{\mu\nu}\tau^2
$$
with $\tau_{\mu\nu}= (\rho +p)u_\mu u_\nu -pg_{\mu\nu}$
we obtain
\be\label{e:brcos}
H^2 = \frac{\kappa_4}{3}\rho\left(1+\frac{\rho}{2\lambda}\right)
 + \frac{C}{a^4} + \frac{\Lambda_4}{3} + \frac{K}{a^2}~,
\ee
where $H$ is the Hubble parameter. In this section we denote the
derivative with respect to \textit{cosmological time} $t$ determined by
$dt=ad\eta$ by an over-dot, so that $H=\dot a/a$ and $\HH=\dot a$. 

The term $\rho^2/(2\la)$ in Eq.~(\ref{e:brcos}) is a correction to the 
Friedmann equation which is important only at high energies and
$\rho^E= C/a^4$, comes from the Weyl tensor. It is called 'Weyl
radiation' since it scales like cosmic radiation, $\propto a^{-4}$.
Observations (nucleosynthesis) tell us that latest at the temperature $T
\sim 1$MeV these additional terms should be unimportant. More
precisely, $\left.\rho^E/\rho\right|_{(nuc)} \lsim 0.1$ and $\la\gsim
(1$MeV$)^4$. For the  5d Planck mass this implies $M_5
\gsim 3\times 10^4$GeV.
The conservation equation of 4-dimensional cosmology remains unchanged,
\be
\dot\rho = -3(\rho + p)\frac{\dot a}{a}~.
\ee
Solutions for $C= K = \La_4 = 0$ are readily found. If the equation
of state is given by $p = w\rho$ we find
\be
   a = a_0\left[t(t+t_\lambda)
     \right]^{\frac{1}{3(1+w)}}    ~,\quad t_{\lambda} =
\frac{M_4}{\sqrt{3\pi\lambda}(3+w)}
< 1{\rm sec}~,
\ee
where we have used $\la>1($Mev$)^4$ for the inequality.
This is to be compared with the usual 4-dimensional behavior from
general relativity (GR). There we have
$ a = a_0t^{2/3(1+w)}$, which corresponds to the above result in the
limit $\la\rightarrow \infty$. Especially
interesting is also the case of de Sitter inflation where
$p = -\rho$ and hence $\rho=$constant, so that $a = a_0e^{Ht}$~, with
\be\label{e:braneH}
 H = \sqrt{\kappa_4\frac{\rho}{3}\left(1+ \frac{\rho}{2\lambda}\right)}
> \sqrt{\frac{\kappa_4\rho}{3}}= H_{GR}~.
\ee
At low energies we recover the usual Friedmann equation while at
high energies, $\rho \ge \la$, the expansion law differs. During
a radiation epoch, $p=\rho/3$ with $\rho\gg \la$ we have
$a\propto t^{1/4}$ (instead of the usual GR behavior, $a\propto t^{1/2}$).
This comes from the fact that in braneworlds at high energies
$H\propto \rho$, where as in 4d GR we have $H\propto\sqrt{\rho}$.
If $\la$ is given by the electroweak scale, $\la \geq 1$TeV$^4$, the observed
low energy cosmology, like nucleosynthesis which starts after
$\rho \sim 1$MeV$^4$ is not affected (if $C$ is sufficiently small).

However, perturbations will carry 5d effects which should in principle
be observable in the fluctuation spectrum of the cosmic microwave
background radiation (CMB) and in the large scale distribution of matter.
These effects are still under investigation. In Section~\ref{sec:pert5}
 I shall  present some preliminary partial results.

\subsection{Brane inflation}
We now want to study scalar field inflation in braneworlds.
Since energy momentum conservation is still valid, the scalar field
evolution equation is also not modified,
$$ \ddot\vph +3H\dot\vph + W'(\vph) = 0~. $$
In 4d GR, the condition for inflation,  $\ddot a>0$  , is
equivalent to $\dot\vph^2 < V$
which corresponds to the violation of the strong energy condition
$$ p = \frac{1}{2}\dot\vph^2 -W < -\frac{1}{3}\rho = -\frac{1}{3}\left(
  \frac{1}{2}\dot\vph^2 +W \right) ~,\quad w= \frac{p}{\rho} < -\frac{1}{3}~,
  \quad \mbox{or}\quad \dot\vph^2 < W~.
$$
The braneworld Friedmann equation (\ref{e:braneH}) leads to a stronger
condition on $w$ for accelerated expansion. For branewords 
$0<\frac{\ddot a}{a} = \dot H  + H^2$ requires
\be\label{e:infcond}
w< -\frac{1}{3}\left[\frac{1+ 2\rho/\lambda}{1+ \rho/\lambda}
\right]~,\quad  \mbox{or}\quad \dot\vph^2 < W
+\left[\frac{\frac{1}{2}\dot\vph^2 
  + W}{\lambda} \left(\frac{5}{4}\dot\vph^2 -
  \frac{1}{2}V \right) \right]
\ee
for inflation to happen. This becomes the usual $w < -1/3$ at low
energy, $\rho<<\la$, but turns into $w < -2/3$ at high energy.

If the slow roll approximation ($\dot\vph^2 \ll W$) is satisfied
we have
$$
H^2 \simeq \frac{\kappa_4}{3}W\left[1 + \frac{W}{2\lambda}\right]
 ~ , \quad \dot\vph \simeq -\frac{W'}{3H}~.
$$
Since the Hubble rate is increased, slow roll is maintained
\textit{longer} than in usual 4d inflation. Correspondingly, the slow roll
parameters are reduced,
\be\label{e:slowroll}
\epsilon_1 \equiv -\frac{\dot H}{H} = \frac{M_4^2}{16\pi}\left(
\frac{W'}{W}\right)^2\left[\frac{1+W/\lambda}
{(1+W/2\lambda)^2}\right] ~, \quad
\ep_2 \equiv -\frac{\ddot\vph}{\dot\vph H} = \frac{M_4^2}{8\pi}\left(
\frac{W''}{W}\right)\left[\frac{1}{(1+W/2\lambda)}\right]~.
\ee
In the high energy regime, $V \gg \la$ they are reduced by factors
$4\la/V$ and $2\la/V$, respectively. Hence, the universe may be inflating
only because it is in the high energy regime and turn into
kinetic dominated expansion, $w\simeq 1$, as soon as $V< \la$. Such
models are constrained since they induce a blue spectrum of gravity
waves~\cite{braneGW}.

In standard 4d GR the perturbation spectrum induced by inflation
is well known and the scalar spectrum agrees extremely well with
the observed anisotropies in the CMB. This will most probably
lead to the most stringent constraints for braneworlds, which however
have not yet been explored in full generality so far. This is still a
very active field of research.

\subsection{Observable consequences from braneworld cosmology}\label{sec:pert5}
So far, we have seen that it is conceivable that our Universe is a
3-brane. At least at low energy and disregarding perturbations, we
cannot distinguish cosmological evolution due to 4d Einstein equations
or the (so different!) brane gravity equations. We finally want to
study ways to discover whether the braneworld idea is realized in nature.
 Are there tell-tale observational signatures
 which would betray whether we live on a brane?

 As we have seen, the gravitational equations for braneworlds
 differ significantly from Einstein gravity. However, at low
 energy the Friedmann equations for a homogeneous and isotropic
 Universe are recovered. Hence there are two regimes in which deviations
 from Einstein gravity will be found:
 \begin{itemize}
 \item {\bf  At high energy.~} This is especially interesting for the
 generation of inflationary perturbations which may be affected by
 high energy braneworld behavior. As long as we restrict ourselves
  on background effects  the calculations are relatively
 straight forward and well under control.
 \item {\bf In the perturbations.~} Cosmological brane perturbation
 theory is not well under control and still a subject of active
 research. One main point are bulk  perturbations which can affect
 the brane and act there like 'sources'. On the other hand,
 gravity wave perturbations generated on the brane can be
 emitted into the bulk.
 \end{itemize}
Here we give examples of both aspects, how effects from braneworlds can enter
cosmological perturbations, but we are by no means exhaustive (more
details and especially references can be found in~\cite{RoyRev}).

\begin{figure}[ht]
\centering
\includegraphics[width=6.5cm]{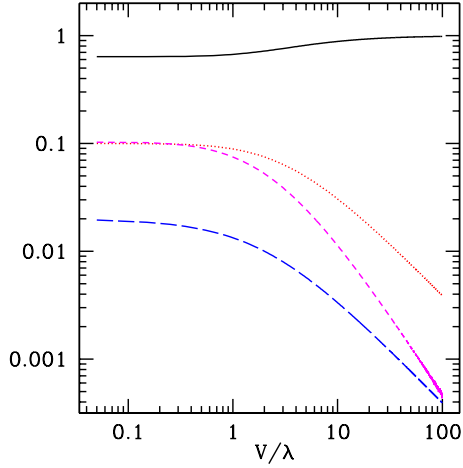}
\caption{We show the slow roll parameters $\ep_1$ (dotted, red) and
  $\ep_2$ (long   dashed, blue) as function of the energy scale of
  inflation, $V/\la$. The spectral index $n$ (solid, black) and the
  tensor to scalar ratio $R$ (dashed, magenta) are also indicated. For
  $W/\la\lsim 1$ the slow roll parameters stay nearly constant and
  correspond to their initial values which are chosen $\ep_1(0) = 0.1$
  and  $\ep_2(0) = 0.02$.}
\label{f:slowroll}
\end{figure}

Let us first consider the high energy universe. During inflation
scalar and tensor perturbations are generated. The spectrum of
 scalar perturbations is $|\Psi|^2k^3 = A_S^2 k^{2q-2} =
  A_S^2 (k/H_0)^{n_s -1}$. The slow roll approximation for braneworlds
  gives~\cite{RoyRev}
\be
A_S^2 \simeq  \frac{512\pi}{75M_4^6}\frac{W^3}{W'^2}
\left[\frac{2\la +W}{2\la}\right]^3 ~, \quad n_S = 1-4\ep_1 -2\ep_2~.
\ee
Similarly, for tensor perturbations, $|H|^2k^3 = A_T^2k^n_T$ one obtains
in the slow roll approximation~\cite{LiSm}
\bea
A_T^2 &\simeq&  \frac{8W}{75M_4^2}F^2(H/\mu)
~, \quad n_T = -2\ep_1 ~,
\quad \mbox{ where } \quad 
F(x) = \left[\sqrt{1+x^2} -x^2\sinh^{-1}(1/x)\right]^{-1/2} \\
\mu &=& \sqrt{\frac{4\pi}{3}}\frac{\sqrt{\la}}{M_4}  \qquad
 \mbox{and } \quad x = H/\mu \simeq \left(\frac{W}{\la}\right)^{-1/2}
\eea
Combining these slow roll equations, at low energy one obtains the same
consistency relation as for ordinary inflation,
\be
\frac{A_T^2}{A_S^2} \simeq \frac{M_4^2W'^2}{16\pi W^2}
    =\ep_1  = -n_T/2~.
\ee
At higher energies, however the relation is different.
Furthermore, the tensor to scalar ratio $R= (A_T/A_S)^2$ and the
spectral indices, both depend on the energy scale $W$.
In Fig.~\ref{f:slowroll} the behavior of the different quantities is
indicated as function of the energy. Of course also the amplitude of
the perturbations strongly depends on the parameter $W/M_4^4$.

\begin{figure}[ht]
\centering
\includegraphics[width=6.5cm]{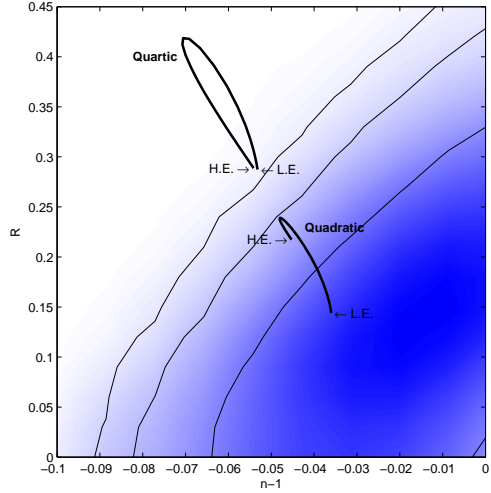}
\caption{
The dependence on the brane tension $\la$ of the parameters $(R,n_s)$
is shown for quartic and quadratic inflation. The 1-, 2- and
3-$\sigma$ contours from the WMAP experiment are also indicated
(from \protect\cite{LiSm}).}
\label{f:braneinf}
\end{figure}

In Fig.~\ref{f:braneinf}  two models are shown, quartic 
inflation with $W= \al\vph^4$ and quadratic inflation with
$W =m^2\vph^2$. The lines of these models in the $(R,n_s)$ plane
for varying $\la$ are drawn. The parapeters $m$ respectively $\al$ are
chosen such that the scalar amplitude is compatible with the
measured value, $A_S^2 \simeq 10^{-10}$ for each brane tension $\la$. The
observational constraints from WMAP data~\cite{WMAP} are also 
indicated. It is clear, that quartic braneworld inflation fares even
worse than ordinary quartic inflation. It is virtually excluded. Also
quadratic inflation with strong braneworld effects fits the data
somewhat less well than usual quadratic inflation, since it predicts
too strong tensor contributions. But clearly, in lack of a concrete
model of inflation (e.g., a given potential) there is little which can
be said.

Discussing the effects on perturbations from braneworlds is opening
Pandora's box. There is a plethora of new phenomena some of which we
don't even know the sign. For example: during
ordinary inflation, gravitational waves are generated. For a given
inflationary potential, their amplitude can be calculated accurately.
However, in the braneworld context, a fraction of these waves will
be radiated into the bulk and thereby reduce the gravity wave
amplitude. On the other hand, there is also gravity wave generation
in the bulk, and some of these accumulate on the brane, increasing the
amplitude of gravity waves on the brane. Therefore, depending on the
precise realization, even the sign of the braneworld effect on a
gravity wave background is unknown.

\begin{figure}[ht]
\centering
\includegraphics[width=10cm]{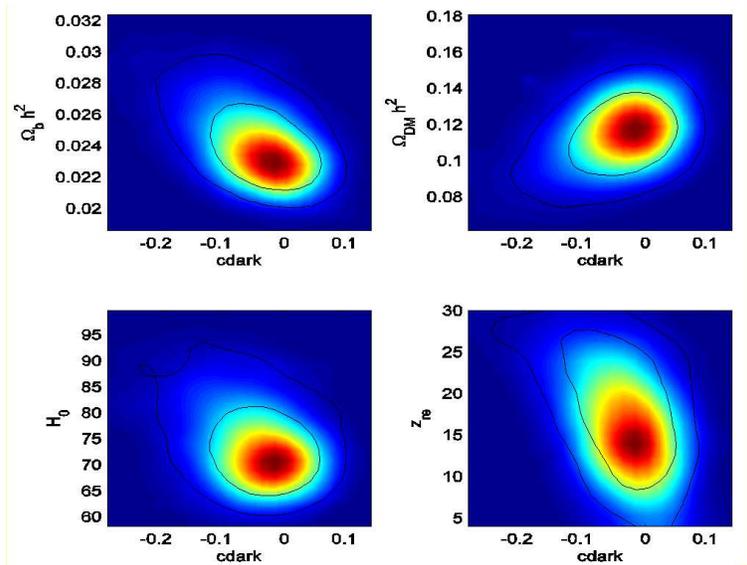}
\caption{
Confidence plots from the CMB data for the amplitude $C_{\rm dark}$ and
other cosmological parameters. The line contours indicate 2- and 3-$\sigma$.
Clearly, a Weyl contribution to the perturbations of more than about 10\% is
strongly disfavored by the data.}
\label{f:WMAPbrane}
\end{figure}

For a more concrete example, let us concentrate on scalar perturbations.
We just take into account, that on the perturbative level the Weyl
tensor $E_{\mu\nu}$ can no longer be neglected. Its energy density perturbation
$\de\rho_E$ acts like a radiation perturbation. In addition, however
it can have an arbitrary amount of anisotropic stress, $\Pi_E$. The
latter induces a difference 
between the two Bardeen potentials, $\Psi-\Phi \propto \Pi_E$. This affects
mainly the Sachs--Wolfe term in the CMB fluctuations, hence the low
multi-poles up to roughly the first peak.
\begin{figure}[ht]
\centering
\includegraphics[width=9cm]{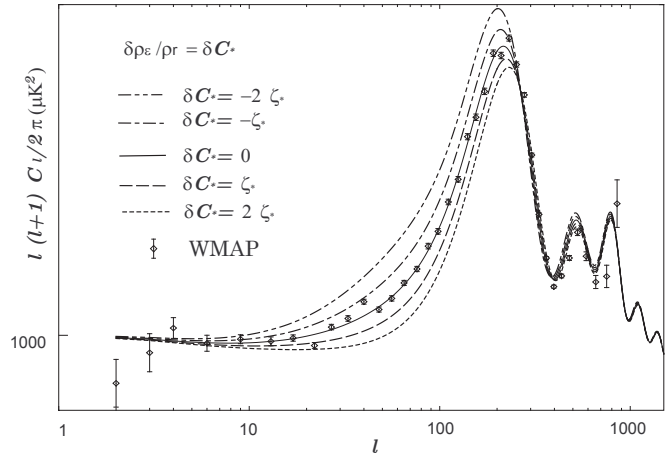}
\caption{
The dependence of the CMB power spectrum on the amplitude of
the Weyl perturbation, $C_{\rm dark}$ for a fixed set of other
cosmological parameters corresponding to the concordance model
(from \protect\cite{Koy}).}
\label{f:CMBbrane}
\end{figure}
In Fig.~\ref{f:CMBbrane} we show the effect of a Weyl perturbation as function
of an amplitude parameter
\be
C_{\rm dark} \equiv  \frac{\de\rho_E/\rho_r}{4\zeta_m}~.
\ee
Here $\zeta_m$ is the $\zeta$ variable defined in Eq.~(\ref{z}), due
to ordinary matter (without the Weyl component). The anisotropic
stress $\Pi_E$ , on large scales, $\ell\ll1/k$, can  be determined
as function of $C_{\rm dark}$ and $\zeta_m$. Confidence plots for the amplitude
$C_{\rm dark}$ and several other cosmological parameters from the WMAP data are
shown in Fig.~\ref{f:WMAPbrane}.

There are many more effects which may come from perturbations in the bulk and
the different perturbation equations on the brane which are presently
under study. A systematic investigation is still lacking.

\section{Conclusions}\label{sec:con}
In these lectures we have studied the possibility that our Universe may
represent a 3-brane in  a higher dimensional space. This idea is
motivated by string theory. We have especially investigated the case
of one large extra dimension where the brane gravitational equations
can be obtained purely from the bulk equations. Even though the
resulting gravity on the brane differs strongly from Einstein gravity,
we have seen that for an Anti--de Sitter bulk, Newton's law is
recovered at large distances and the Friedmann equations for the
evolution of the Universe are obtained at low energy.

It is, however by no means clear, to which extent the different
gravitational equations will spoil the successes of cosmological
perturbation theory.  This is still an open question and its
answer will be crucial for braneworlds.
\vspace{0.4cm}

\noindent
{\bf Acknowledgment:~~} I am most grateful to the organizers for inviting
me to this successful, traditional school in such a beautiful
environment. I warmly thank also the students for their interest and
active participation which was a most stimulating experience
for me. I thank Marcus Ruser for carefully reading the manuscript.


\begin{thebibliography}{99}

\bibitem{Pol}J.\ Polchinski, \textit{String theory.\ Vol.\ I:
An Introduction to the  Bosonic string}, Cambridge University Press (1998);\\
 \textit{String theory.\ Vol.\ II:
  Superstring theory and beyond}, Cambridge University Press (1998).
\bibitem{micGra} J. Chiaverini \etal Phys. Rev. Lett. {\bf 90} 151101
  (2003);\\
J.C. Long \etal, Nature {\bf 421} 922 (2003).
\bibitem{ADD}N. Arkani-Hamed, S. Dimopoulos and G. Dvali,Phys. Lett. {\bf
  B429}, 263 (1998) [e-print: hep-ph/9803315].
\bibitem{bounds}N. Arkani-Hamed, S. Dimopoulos and G. Dvali,
  Phys. Rev. {\bf D59}, 086004 (1999)  [e-print: hep-ph/9807344].
\bibitem{Wein}S. Weinberg, \textit{Quantum  Theory of Fields, Vol.~I},
  Cambridge University Press (1995).
\bibitem{KK}T. Kaluza, Sitzungsber. Preuss. Akad. Wiss. Berlin, 966
            (1921);\\
            O. Klein, Z.Phys. {\bf 37}, 895 (1926).
\bibitem{bipul}C.M. Will, Living Rev. 
        Relativity {\bf 4}, 4 (2001): cited on March 2005, 
        ~~~ http://wwwlivingreviews.org/Articles/Volume4/ 
\bibitem{DK}R. Durrer and P. Kocian, Class. Quant. Grav. {\bf 21} ,
         2127 (2004) [e-print:hep-th/0305181].
\bibitem{GW} W. Goldberger and M. Wise, Phys. Rev. Lett {\bf 83}, 4922
  (1999) [e-print: hep-ph/9907447].
\bibitem{HOWI} P. Horawa and E. Witten, Nucl. Phys. {\bf B60}, 506 (1996).
\bibitem{Is}W. Israel, Nuovo Cimento {\bf 44B}, 463 (1966).
\bibitem{La}C. Lanczos, {Ann. Phys.} {\bf 74}, 518 (1924).
\bibitem{Da}G. Darmois, M{\'e}morial des sciences math{\'e}matiques, fasicule
  25 chap. 5, Gauthier-Villars (Paris, 1927).
\bibitem{Mi}C. Misner, {K.~S.} Thorne, and J.~A.
  Wheeler, {\em Gravitation}
  (W.H. Freeman and Compagny, San Francisco, USA 1973).
\bibitem{Shiru}T. Shirumizu, K. Maeda and M. Sasaki, Phys. Rev. {\bf
 D62}, 024012 (2000)  [e-print gr-qc/9910076].
\bibitem{RoyRev} R.\ Maartens, Living Rev.\ Relativity {\bf 7}  (2004).\\
 Cited on 11/17/04,~~~ http://www.livingreviews.org/lrr-2004-7
\bibitem{RS1}L. Randall and R. Sundrum, Phys. Rev. Lett. {\bf 83}, 3370
  (1999) [e-print: hep-ph/9905221].
\bibitem{RS2}L. Randall and R. Sundrum, Phys. Rev. Lett.  {\bf 83}, 4690
  (1999)  [e-print: hep-ph/9906064].
\bibitem{Csaki:2000fc}C. Csaki, J. Erlich, T. Hollowood, and
  Y. Shirman, Nucl. Phys. {\bf B591}, 309 (2000)  [e-print
    hep-th/0001033].
\bibitem{Bozza:2001xt} V. Bozza, M. Gasperini, and G.  Veneziano,
  Nucl. Phys. {\bf B619}, 191 (2001)  [e-print hep-th/0004067].
\bibitem{CD4}C. Cartier and R. Durrer,  [e-print
    hep-th/0408287].
\bibitem{Mukohyama:2000ui}S. Mukohyama,
{Phys. Rev.} {\bf D62}, 084015 (2000) [e-print hep-th/0004067].
\bibitem{Bridgman:2000ih}H. Bridgman, K. Malik,
  and D. Wands, Phys. Rev. {\bf D63},
 084012 (2001) [e-print hep-th/0010133].
\bibitem{Ringeval:2003na}C. Ringeval, T. Boehm and R. Durrer, [e-print
  hep-th/0307100].
\bibitem{Csaki:1999jh}C. Csaki, M. Graesser, C.   Kolda, and
  J. Terning, Phys. Lett. {\bf B462}, 34 (1999) [e-print hep-ph/9906513].
\bibitem{CoHi}R. Courant and D. Hilbert, {\em Methods in Mathematical
  Physics Vol. I}, J. Wiley \& Sons (1989).
\bibitem{TG}J. Garriga and T. Tanaka, Phys. Rev. Lett. {\bf 84}, 2778
  (2000) [e-print hep-th/9911055].
\bibitem{Rfund}
        R. Durrer, { Fund. of Cosmic Physics} {\bf 15}, 209 (1994)
        [e-print astro-ph/9311041]. 
\bibitem{Dod}S. Dodelson,  \textit{Modern Cosmology}, Academic Press
  (London, 2003).
\bibitem{peebles}
    P.J.E. Peebles, { \em Principles of Physical Cosmology},
    Princeton University Press (1993).
\bibitem{Ellis} M. Bruni, P. Dunsby and G. Ellis, { Astrophys. J.}
    {\bf 395}, 34 (1992).
\bibitem{Mukhanov:1992tc}V. F. Mukhanov, H. A. Feldman and
        R. H. Brandenberger, Phys. Rept. {\bf 215}, 203 (1992).
\bibitem{Lif46}E. Lifshitz, JETP {\bf 10}, 116 (1946).
\bibitem{HZ}
    E. Harrison, { Phys. Rev.} {\bf D1} 2726 (1970);\\
     Ya. B. Zel'dovich, { Mont. Not. R. Astr. Soc.}
        {\bf 160}, P1 (1972).
\bibitem{LiLy}A. Liddle and D.Lyth,  \textit{Cosmological Inflation
  and Large Scale Structure}, Cambridge University Press (2000).
\bibitem{Linde}A. Linde, \textit{Particle Physics and Inflationary
  Cosmology}, Harwood Academics (Chur, 1990).
\bibitem{myCMB}R. Durrer, {\em ``Cosmological Perturbation Theory''},
  in: The Physics of the Early Universe, Lecture Notes in Physics, Volume 653,
      edt. E. Papantonopoulos, Springer Verlag (2004)
      [e-print astro-ph/0402129].
\bibitem{SW}
    R.K. Sachs and A.M. Wolfe, { Astrophys. J.} {\bf 147}, 73 (1967).
\bibitem{DMR}
    G.F. Smoot et al., { Astrophys. J.} {\bf 396}, L1 (1992).
\bibitem{WMAP} C.L. Bennett et al., Astrophys. J. Suppl. {\bf 148}
        (2003) 1;\\
        G. Hinshaw et al.,   Astrophys. J. Suppl. {\bf 148}
        (2003) 135;\\
         D.N. Spergel et al.,  Astrophys. J. Suppl. {\bf 148}
        (2003) 175;\\
        and the  other publications in the same issue of  Astrophys. J. Suppl.
\bibitem{report}R. Durrer, M. Kunz and A. Melchiorri, Phys. Rept. {\bf
        364}, 1 (2002) [e-print: astro-ph/0110348].
\bibitem{dgsv} R. Durrer {\it et al.}, {Phys. Rev. D} {\bf 59}, 043511 (1999)
 [e-print gr-qc/9804076].
\bibitem{JL}W. Jones and A. Lasenby, Living Reviews of Relativity
  {\bf 1}, 11 (1998). [Online article: cited on 1 March 2005,
http://www.livingreviews.org/lrr-1998-11]
\bibitem{braneGW} V. Sahni, M. Sami and T. Souradeep, Phys. Rev. {\bf
  D65}, 023518 (2002)  [e-print gr-qc/0105121].

\bibitem{LiSm}A.Liddle and A. Smith, Phys. Rev. {\bf D68}, 061301
  (2003) [e-print astro-ph/0307017].
\bibitem{Koy}K. Koyama, Phys. Rev. Lett. {\bf 91}, 221301
  (2003) [e-print astro-ph/0303108]. 

\end{thebibliography}
\end{document}